\documentstyle[aps,pra]{revtex}
\def\temp{1.34}%
\let\tempp=\relax
\expandafter\ifx\csname psboxversion\endcsname\relax
\else
    \ifdim\temp cm>\psboxversion cm
    \else
      \let\temp=\psboxversion
      \let\tempp= 
    \fi
\fi
\tempp
\let\psboxversion=\temp
\catcode`\@=11
\def\psfortextures{
\def\PSspeci@l##1##2{%
\special{illustration ##1\space scaled ##2}%
}}%
\def\psfordvitops{
\def\PSspeci@l##1##2{%
\special{dvitops: import ##1\space \the\drawingwd \the\drawinght}%
}}%
\def\psfordvips{
\def\PSspeci@l##1##2{%
\d@my=0.1bp \d@mx=\drawingwd \divide\d@mx by\d@my
\includegraphics{##1\space}}}%
\def\psforoztex{
\def\PSspeci@l##1##2{%
\special{##1 \space
      ##2 1000 div dup scale
      \number-\psllx\space \number-\pslly\space translate
}}}%
\def\psfordvitps{
\def\psdimt@n@sp##1{\d@mx=##1\relax\edef\psn@sp{\number\d@mx}}
\def\PSspeci@l##1##2{%
\special{dvitps: Include0 "psfig.psr"}
\psdimt@n@sp{\drawingwd}
\special{dvitps: Literal "\psn@sp\space"}
\psdimt@n@sp{\drawinght}
\special{dvitps: Literal "\psn@sp\space"}
\psdimt@n@sp{\psllx bp}
\special{dvitps: Literal "\psn@sp\space"}
\psdimt@n@sp{\pslly bp}
\special{dvitps: Literal "\psn@sp\space"}
\psdimt@n@sp{\psurx bp}
\special{dvitps: Literal "\psn@sp\space"}
\psdimt@n@sp{\psury bp}
\special{dvitps: Literal "\psn@sp\space startTexFig\space"}
\special{dvitps: Include1 "##1"}
\special{dvitps: Literal "endTexFig\space"}
}}%
\def\psfordvialw{
\def\PSspeci@l##1##2{
\special{language "PostScript",
position = "bottom left",
literal "  \psllx\space \pslly\space translate
  ##2 1000 div dup scale
  -\psllx\space -\pslly\space translate",
include "##1"}
}}%
\def\psforptips{
\def\PSspeci@l##1##2{{
\d@mx=\psurx bp
\advance \d@mx by -\psllx bp
\divide \d@mx by 1000\multiply\d@mx by \xscale
\incm{\d@mx}
\let\tmpx\dimincm
\d@my=\psury bp
\advance \d@my by -\pslly bp
\divide \d@my by 1000\multiply\d@my by \xscale
\incm{\d@my}
\let\tmpy\dimincm
\d@mx=-\psllx bp
\divide \d@mx by 1000\multiply\d@mx by \xscale
\d@my=-\pslly bp
\divide \d@my by 1000\multiply\d@my by \xscale
\at(\d@mx;\d@my){\special{ps:##1 x=\tmpx, y=\tmpy}}
}}}%
\def\psonlyboxes{
\def\PSspeci@l##1##2{%
\at(0cm;0cm){\boxit{\vbox to\drawinght
  {\vss\hbox to\drawingwd{\at(0cm;0cm){\hbox{({\tt##1})}}\hss}}}}
}}%
\def\psloc@lerr#1{%
\let\savedPSspeci@l=\PSspeci@l%
\def\PSspeci@l##1##2{%
\at(0cm;0cm){\boxit{\vbox to\drawinght
  {\vss\hbox to\drawingwd{\at(0cm;0cm){\hbox{({\tt##1}) #1}}\hss}}}}
\let\PSspeci@l=\savedPSspeci@l
}}%
\newread\pst@mpin
\newdimen\drawinght\newdimen\drawingwd
\newdimen\psxoffset\newdimen\psyoffset
\newbox\drawingBox
\newcount\xscale \newcount\yscale \newdimen\pscm\pscm=1cm
\newdimen\d@mx \newdimen\d@my
\newdimen\pswdincr \newdimen\pshtincr
\let\ps@nnotation=\relax
{\catcode`\|=0 |catcode`|\=12 |catcode`|
|catcode`#=12 |catcode`*=14
|xdef|backslashother{\}*
|xdef|percentother{
|xdef|tildeother{~}*
|xdef|sharpother{#}*
}%
\def\R@moveMeaningHeader#1:->{}%
\def\uncatcode#1{%
\edef#1{\expandafter\R@moveMeaningHeader\meaning#1}}%
\def\execute#1{#1}
\def\psm@keother#1{\catcode`#112\relax}
\def\executeinspecs#1{%
\execute{\begingroup\let\do\psm@keother\dospecials\catcode`\^^M=9#1\endgroup}}%
\def\@mpty{}%
\def\matchexpin#1#2{
  \fi%
  \edef\tmpb{{#2}}%
  \expandafter\makem@tchtmp\tmpb%
  \edef\tmpa{#1}\edef\tmpb{#2}%
  \expandafter\expandafter\expandafter\m@tchtmp\expandafter\tmpa\tmpb\endm@tch%
  \if\match%
}%
\def\matchin#1#2{%
  \fi%
  \makem@tchtmp{#2}%
  \m@tchtmp#1#2\endm@tch%
  \if\match%
}%
\def\makem@tchtmp#1{\def\m@tchtmp##1#1##2\endm@tch{%
  \def\tmpa{##1}\def\tmpb{##2}\let\m@tchtmp=\relax%
  \ifx\tmpb\@mpty\def\match{YN}%
  \else\def\match{YY}\fi%
}}%
\def\incm#1{{\psxoffset=1cm\d@my=#1
 \d@mx=\d@my
  \divide\d@mx by \psxoffset
  \xdef\dimincm{\number\d@mx.}
  \advance\d@my by -\number\d@mx cm
  \multiply\d@my by 100
 \d@mx=\d@my
  \divide\d@mx by \psxoffset
  \edef\dimincm{\dimincm\number\d@mx}
  \advance\d@my by -\number\d@mx cm
  \multiply\d@my by 100
 \d@mx=\d@my
  \divide\d@mx by \psxoffset
  \xdef\dimincm{\dimincm\number\d@mx}
}}%
\newif\ifNotB@undingBox
\newhelp\PShelp{Proceed: you'll have a 5cm square blank box instead of
your graphics (Jean Orloff).}%
\def\s@tsize#1 #2 #3 #4\@ndsize{
  \def\psllx{#1}\def\pslly{#2}%
  \def\psurx{#3}\def\psury{#4}
  \ifx\psurx\@mpty\NotB@undingBoxtrue
  \else
    \drawinght=#4bp\advance\drawinght by-#2bp
    \drawingwd=#3bp\advance\drawingwd by-#1bp
  \fi
  }%
\def\sc@nBBline#1:#2\@ndBBline{\edef\p@rameter{#1}\edef\v@lue{#2}}%
\def\g@bblefirstblank#1#2:{\ifx#1 \else#1\fi#2}%
{\catcode`\%=12
\xdef\B@undingBox{
\def\ReadPSize#1{
 \readfilename#1\relax
 \let\PSfilename=\lastreadfilename
 \openin\pst@mpin=#1\relax
 \ifeof\pst@mpin \errhelp=\PShelp
   \errmessage{I haven't found your postscript file (\PSfilename)}%
   \psloc@lerr{was not found}%
   \s@tsize 0 0 142 142\@ndsize
   \closein\pst@mpin
 \else
   \if\matchexpin{\GlobalInputList}{, \lastreadfilename}%
   \else\xdef\GlobalInputList{\GlobalInputList, \lastreadfilename}%
     \immediate\write\psbj@inaux{\lastreadfilename,}%
   \fi%
   \loop
     \executeinspecs{\catcode`\ =10\global\read\pst@mpin to\n@xtline}%
     \ifeof\pst@mpin
       \errhelp=\PShelp
       \errmessage{(\PSfilename) is not an Encapsulated PostScript File:
           I could not find any \B@undingBox: line.}%
       \edef\v@lue{0 0 142 142:}%
       \psloc@lerr{is not an EPSFile}%
       \NotB@undingBoxfalse
     \else
       \expandafter\sc@nBBline\n@xtline:\@ndBBline
       \ifx\p@rameter\B@undingBox\NotB@undingBoxfalse
         \edef\t@mp{%
           \expandafter\g@bblefirstblank\v@lue\space\space\space}%
         \expandafter\s@tsize\t@mp\@ndsize
       \else\NotB@undingBoxtrue
       \fi
     \fi
   \ifNotB@undingBox\repeat
   \closein\pst@mpin
 \fi
\message{#1}%
}%
\def\psboxto(#1;#2)#3{\vbox{%
   \ReadPSize{#3}%
   \advance\pswdincr by \drawingwd
   \advance\pshtincr by \drawinght
   \divide\pswdincr by 1000
   \divide\pshtincr by 1000
   \d@mx=#1
   \ifdim\d@mx=0pt\xscale=1000
         \else \xscale=\d@mx \divide \xscale by \pswdincr\fi
   \d@my=#2
   \ifdim\d@my=0pt\yscale=1000
         \else \yscale=\d@my \divide \yscale by \pshtincr\fi
   \ifnum\yscale=1000
         \else\ifnum\xscale=1000\xscale=\yscale
                    \else\ifnum\yscale<\xscale\xscale=\yscale\fi
              \fi
   \fi
   \divide\drawingwd by1000 \multiply\drawingwd by\xscale
   \divide\drawinght by1000 \multiply\drawinght by\xscale
   \divide\psxoffset by1000 \multiply\psxoffset by\xscale
   \divide\psyoffset by1000 \multiply\psyoffset by\xscale
   \global\divide\pscm by 1000
   \global\multiply\pscm by\xscale
   \multiply\pswdincr by\xscale \multiply\pshtincr by\xscale
   \ifdim\d@mx=0pt\d@mx=\pswdincr\fi
   \ifdim\d@my=0pt\d@my=\pshtincr\fi
   \message{scaled \the\xscale}%
 \hbox to\d@mx{\hss\vbox to\d@my{\vss
   \global\setbox\drawingBox=\hbox to 0pt{\kern\psxoffset\vbox to 0pt{%
      \kern-\psyoffset
      \PSspeci@l{\PSfilename}{\the\xscale}%
      \vss}\hss\ps@nnotation}%
   \global\wd\drawingBox=\the\pswdincr
   \global\ht\drawingBox=\the\pshtincr
   \global\drawingwd=\pswdincr
   \global\drawinght=\pshtincr
   \baselineskip=0pt
   \copy\drawingBox
 \vss}\hss}%
  \global\psxoffset=0pt
  \global\psyoffset=0pt
  \global\pswdincr=0pt
  \global\pshtincr=0pt 
  \global\pscm=1cm 
}}%
\def\psboxscaled#1#2{\vbox{%
  \ReadPSize{#2}%
  \xscale=#1
  \message{scaled \the\xscale}%
  \divide\pswdincr by 1000 \multiply\pswdincr by \xscale
  \divide\pshtincr by 1000 \multiply\pshtincr by \xscale
  \divide\psxoffset by1000 \multiply\psxoffset by\xscale
  \divide\psyoffset by1000 \multiply\psyoffset by\xscale
  \divide\drawingwd by1000 \multiply\drawingwd by\xscale
  \divide\drawinght by1000 \multiply\drawinght by\xscale
  \global\divide\pscm by 1000
  \global\multiply\pscm by\xscale
  \global\setbox\drawingBox=\hbox to 0pt{\kern\psxoffset\vbox to 0pt{%
     \kern-\psyoffset
     \PSspeci@l{\PSfilename}{\the\xscale}%
     \vss}\hss\ps@nnotation}%
  \advance\pswdincr by \drawingwd
  \advance\pshtincr by \drawinght
  \global\wd\drawingBox=\the\pswdincr
  \global\ht\drawingBox=\the\pshtincr
  \global\drawingwd=\pswdincr
  \global\drawinght=\pshtincr
  \baselineskip=0pt
  \copy\drawingBox
  \global\psxoffset=0pt
  \global\psyoffset=0pt
  \global\pswdincr=0pt
  \global\pshtincr=0pt 
  \global\pscm=1cm
}}%
\def\psbox#1{\psboxscaled{1000}{#1}}%
\newif\ifn@teof\n@teoftrue
\newif\ifc@ntrolline
\newif\ifmatch
\newread\j@insplitin
\newwrite\j@insplitout
\newwrite\psbj@inaux
\immediate\openout\psbj@inaux=psbjoin.aux
\immediate\write\psbj@inaux{\string\joinfiles}%
\immediate\write\psbj@inaux{\jobname,}%
\def\toother#1{\ifcat\relax#1\else\expandafter%
  \toother@ux\meaning#1\endtoother@ux\fi}%
\def\toother@ux#1 #2#3\endtoother@ux{\def\tmp{#3}%
  \ifx\tmp\@mpty\def\tmp{#2}\let\next=\relax%
  \else\def\next{\toother@ux#2#3\endtoother@ux}\fi%
\next}%
\let\readfilenamehook=\relax
\def\re@d{\expandafter\re@daux}
\def\re@daux{\futurelet\nextchar\stopre@dtest}%
\def\re@dnext{\xdef\lastreadfilename{\lastreadfilename\nextchar}%
  \afterassignment\re@d\let\nextchar}%
\def\stopre@d{\egroup\readfilenamehook}%
\def\stopre@dtest{%
  \ifcat\nextchar\relax\let\nextread\stopre@d
  \else
    \ifcat\nextchar\space\def\nextread{%
      \afterassignment\stopre@d\chardef\nextchar=`}%
    \else\let\nextread=\re@dnext
      \toother\nextchar
      \edef\nextchar{\tmp}%
    \fi
  \fi\nextread}%
\def\readfilename{\bgroup%
  \let\\=\backslashother \let\%=\percentother \let\~=\tildeother
  \let\#=\sharpother \xdef\lastreadfilename{}%
  \re@d}%
\xdef\GlobalInputList{\jobname}%
\def\psnewinput{%
  \def\readfilenamehook{
    \if\matchexpin{\GlobalInputList}{, \lastreadfilename}%
    \else\xdef\GlobalInputList{\GlobalInputList, \lastreadfilename}%
      \immediate\write\psbj@inaux{\lastreadfilename,}%
    \fi%
    \ps@ldinput\lastreadfilename\relax%
    \let\readfilenamehook=\relax%
  }\readfilename%
}%
\expandafter\ifx\csname @@input\endcsname\relax    
  \immediate\let\ps@ldinput=\input\def\input{\psnewinput}%
\else
  \immediate\let\ps@ldinput=\@@input
  \def\@@input{\psnewinput}%
\fi%
\def\nowarnopenout{%
 \def\warnopenout##1##2{%
   \readfilename##2\relax
   \message{\lastreadfilename}%
   \immediate\openout##1=\lastreadfilename\relax}}%
\def\warnopenout#1#2{%
 \readfilename#2\relax
 \def\t@mp{TrashMe,psbjoin.aux,psbjoint.tex,}\uncatcode\t@mp
 \if\matchexpin{\t@mp}{\lastreadfilename,}%
 \else
   \immediate\openin\pst@mpin=\lastreadfilename\relax
   \ifeof\pst@mpin
     \else
     \errhelp{If the content of this file is so precious to you, abort (ie
press x or e) and rename it before retrying.}%
     \errmessage{I'm just about to replace your file named \lastreadfilename}%
   \fi
   \immediate\closein\pst@mpin
 \fi
 \message{\lastreadfilename}%
 \immediate\openout#1=\lastreadfilename\relax}%
{\catcode`\%=12\catcode`\*=14
\gdef\splitfile#1{*
 \readfilename#1\relax
 \immediate\openin\j@insplitin=\lastreadfilename\relax
 \ifeof\j@insplitin
   \message{! I couldn't find and split \lastreadfilename!}*
 \else
   \immediate\openout\j@insplitout=TrashMe
   \message{< Splitting \lastreadfilename\space into}*
   \loop
     \ifeof\j@insplitin
       \immediate\closein\j@insplitin\n@teoffalse
     \else
       \n@teoftrue
       \executeinspecs{\global\read\j@insplitin to\spl@tinline\expandafter
         \ch@ckbeginnewfile\spl@tinline
       \ifc@ntrolline
       \else
         \toks0=\expandafter{\spl@tinline}*
         \immediate\write\j@insplitout{\the\toks0}*
       \fi
     \fi
   \ifn@teof\repeat
   \immediate\closeout\j@insplitout
 \fi\message{>}*
}*
\gdef\ch@ckbeginnewfile#1
 \def\t@mp{#1}*
 \ifx\@mpty\t@mp
   \def\t@mp{#3}*
   \ifx\@mpty\t@mp
     \global\c@ntrollinefalse
   \else
     \immediate\closeout\j@insplitout
     \warnopenout\j@insplitout{#2}*
     \global\c@ntrollinetrue
   \fi
 \else
   \global\c@ntrollinefalse
 \fi}*
\gdef\joinfiles#1\into#2{*
 \message{< Joining following files into}*
 \warnopenout\j@insplitout{#2}*
 \message{:}*
 {*
 \edef\w@##1{\immediate\write\j@insplitout{##1}}*
\w@{
\w@{
\w@{
\w@{
\w@{
\w@{
\w@{
\w@{
\w@{
\w@{
\w@{\string\input\space psbox.tex}*
\w@{\string\splitfile{\string\jobname}}*
\w@{\string\let\string\autojoin=\string\relax}*
}*
 \expandafter\tre@tfilelist#1, \endtre@t
 \immediate\closeout\j@insplitout
 \message{>}*
}*
\gdef\tre@tfilelist#1, #2\endtre@t{*
 \readfilename#1\relax
 \ifx\@mpty\lastreadfilename
 \else
   \immediate\openin\j@insplitin=\lastreadfilename\relax
   \ifeof\j@insplitin
     \errmessage{I couldn't find file \lastreadfilename}*
   \else
     \message{\lastreadfilename}*
     \immediate\write\j@insplitout{
     \executeinspecs{\global\read\j@insplitin to\oldj@ininline}*
     \loop
       \ifeof\j@insplitin\immediate\closein\j@insplitin\n@teoffalse
       \else\n@teoftrue
         \executeinspecs{\global\read\j@insplitin to\j@ininline}*
         \toks0=\expandafter{\oldj@ininline}*
         \let\oldj@ininline=\j@ininline
         \immediate\write\j@insplitout{\the\toks0}*
       \fi
     \ifn@teof
     \repeat
   \immediate\closein\j@insplitin
   \fi
   \tre@tfilelist#2, \endtre@t
 \fi}*
}%
\def\autojoin{%
 \immediate\write\psbj@inaux{\string\into{psbjoint.tex}}%
 \immediate\closeout\psbj@inaux
 \expandafter\joinfiles\GlobalInputList\into{psbjoint.tex}%
}%
\def\centinsert#1{\midinsert\line{\hss#1\hss}\endinsert}%
\def\psannotate#1#2{\vbox{%
  \def\ps@nnotation{#2\global\let\ps@nnotation=\relax}#1}}%
\def\pscaption#1#2{\vbox{%
   \setbox\drawingBox=#1
   \copy\drawingBox
   \vskip\baselineskip
   \vbox{\hsize=\wd\drawingBox\setbox0=\hbox{#2}%
     \ifdim\wd0>\hsize
       \noindent\unhbox0\tolerance=5000
    \else\centerline{\box0}%
    \fi
}}}%
\def\at(#1;#2)#3{\setbox0=\hbox{#3}\ht0=0pt\dp0=0pt
  \rlap{\kern#1\vbox to0pt{\kern-#2\box0\vss}}}%
\newdimen\gridht \newdimen\gridwd
\def\gridfill(#1;#2){%
  \setbox0=\hbox to 1\pscm
  {\vrule height1\pscm width.4pt\leaders\hrule\hfill}%
  \gridht=#1
  \divide\gridht by \ht0
  \multiply\gridht by \ht0
  \gridwd=#2
  \divide\gridwd by \wd0
  \multiply\gridwd by \wd0
  \advance \gridwd by \wd0
  \vbox to \gridht{\leaders\hbox to\gridwd{\leaders\box0\hfill}\vfill}}%
\def\fillinggrid{\at(0cm;0cm){\vbox{%
  \gridfill(\drawinght;\drawingwd)}}}%
\def\textleftof#1:{%
  \setbox1=#1
  \setbox0=\vbox\bgroup
    \advance\hsize by -\wd1 \advance\hsize by -2em}%
\def\textrightof#1:{%
  \setbox0=#1
  \setbox1=\vbox\bgroup
    \advance\hsize by -\wd0 \advance\hsize by -2em}%
\def\endtext{%
  \egroup
  \hbox to \hsize{\valign{\vfil##\vfil\cr%
\box0\cr%
\noalign{\hss}\box1\cr}}}%
\def\frameit#1#2#3{\hbox{\vrule width#1\vbox{%
  \hrule height#1\vskip#2\hbox{\hskip#2\vbox{#3}\hskip#2}%
        \vskip#2\hrule height#1}\vrule width#1}}%
\def\boxit#1{\frameit{0.4pt}{0pt}{#1}}%
\catcode`\@=12 
 \psfordvips   

\tighten
\begin{document}
\title{Effective Breather Trapping Mechanism for DNA Transcription}
\author{Julian J.-L. Ting}
\address{
Institute f\"ur Festk\"orperforschung,
Forschungszentrum J\"ulich GmbH,
Postfach 1913
D-52428 J\"ulich, Germany
\footnote{
On leave of absence from
Physics Institute, Tsing-hua University, 
Hsin-Chu, Taiwan  30043, Republic of China.
} 
}
\author{Michel Peyrard}   
\address{
Laboratoire de Physique de l'Ecole Normale Sup\'erieure de Lyon,
CNRS URA 1325,
46 all\'ee d'Italie, 69007 Lyon, France.
\footnote{E-mail address: mpeyrard@physique.ens-lyon.fr } \\
}

\date{Phys. Rev. E {\bf 53} 1011-1020 (1996)}
\maketitle
\begin{abstract}
Collective coordinate and direct numerical integration methods are applied to
the analysis of a  one-dimensional DNA model. A modification of the coupling
constant in an extended region is found to be less selective towards the
breather it can trap than an isolated impurity. Therefore it provides a possible
physical mechanism for the effect of an enzyme on DNA transcription.
\\ PACS:  87.10.+e, 03.40.Kf
\end{abstract} 

\section{Introduction}

The first step of the transcription of deoxyribonucleic acid (DNA) is a local
opening of the double helix which extends over about 20 base pairs. Such
local unwindings of the helix can be obtained by heating DNA to about
$70^{\circ}~C$. But in the life of an organism they must occur at
physiological temperature. This is achieved by the action of an
enzyme\cite{CD}. However one may wonder how this can be possible since,
whatever its origin, the local opening requires the breaking of the same
number of hydrogen bonds, hence the same amount of energy, and the enzyme
does not bring in  energy. However, under normal physiological conditions
there are thermal fluctuations along the DNA chain. They can be weakly
localized by nonlinear effects to generate what biologists call the
``breathing of DNA''. But their intensity is not high enough to open the
double helix over many base pairs.  A possible pathway to the opening would
be to collect the thermal energy that is present along the molecule.  This
could be the role of the enzyme. From a physicist point of view, the effect
of an enzyme can be considered as a  perturbation to the DNA lattice.

Recently Forinash {\it et al} considered the interaction between a mass
impurity on a DNA chain,  and thermal nonlinear waves described  as
breathers traveling along the chain.\cite{FPM}  They found that the
impurity is selective toward the breather it can trap. Although this is
a first indication that a defect can contribute to localize energy in a
nonlinear chain, it does not appear to be a good model for the action of
an enzyme because, with such a localized defect, only some predefined
frequencies of the thermal fluctuations would contribute to bring in the
energy. Therefore one may ask whether there exist any other mechanism
more efficient to trap energy.

One learns from biological studies that some proteins, make contact with DNA
at multiple sites\cite{A,EYSB}.  Moreover the transcription enzyme actually
bends DNA toward itself.  It has the effect not only to modify the mass  at
some sites but also to modify the coupling constants along the strands. The
bases which are inside the bend are brought closer to each other while the
ones which are outside are moved farther apart. Although the variation of
the distances between neighboring bases may be rather small, it can have a
large effect because the interaction between bases is due to the overlap of
$\pi$ electrons over the whole surface of the planar bases. We examine in
this paper whether the interaction of the enzyme with more than one site 
might be more efficient for trapping breathers than isolated impurities by
studying the effect of an extended modification of the coupling along the
DNA chain.

The effect of bending and twisting to modify the elasticity of DNA has been
considered previously by Barkley and Zimm\cite{BZ}, 
and by Marko and Siggia\cite{MS} but they did not study the
consequences of base pair opening.  Salerno\cite{S} considered the dynamical
properties of a DNA promoter   which has some similarities with our problem
because we treat here the enzyme as an inhomogeneity due to an external
effect while he considers inhomogeneities from the DNA composition itself. 
However he was interested by kinks while we study breathing modes. In a more
abstract level we are investigating here a nonlinear model, with  an
``extended defect'', and we try to understand the interplay between
nonlinearity and disorder. In the harmonic case, a one dimensional chain with
isolated defects has been considered before by Montroll and Potts\cite{MP}.
However, besides the introduction of nonlinearity, one should also notice
that for the type of extended defect that we consider, there is no
evanescent local mode which would couple to a breather as in the case
considered by Forinash {\it et al}, so that the mechanism for energy
localization must be different.

\section{DNA lattice model}

If one neglects the small longitudinal motion and concentrates on the
stretching of the base pairs, DNA can be described by  a simple
one-dimensional model\cite{PB} which consists of an array of harmonically
coupled particles subjected  to a Morse potential. Such a model is
sufficient to provide a good qualitative description of the thermal
denaturation of the molecule\cite{Dauxois}. If one treats a chain with
inhomogeneous coupling, the equations of motion read
\begin{eqnarray}
\label{discrete}
  m {{\partial^2 Y_n} \over {\partial T^2}} - 
  K_{n+1} ( Y_{n+1} - Y_n ) + K_n ( Y_n - Y_{n-1} ) 
  \nonumber \\
  - 2 D \alpha  e^{- \alpha Y_n} 
  ( e^{- \alpha Y_n} - 1)  = 0 \; ,
\end{eqnarray}
in which $\alpha$ and $D$ are parameters for the Morse potential\cite{M},
which have dimensions of inverse length and energy respectively, and $n$ is
the site index. 
Fig.~\ref{line1} shows the geometry and the coordinate used. It is convenient
for the analytical calculations to transform these equations into a
dimensionless form by defining the following dimensionless variables:

\begin{eqnarray}
y_n &=& \alpha Y_n, \\
t_{~} &=& \sqrt {{D \alpha^2 } \over m} T , \\
k_n &=& {K_n \over {D \alpha^2 }}. 
\end{eqnarray}
The equations become
\begin{eqnarray}
\label{ND}
  {{\partial^2 y_n} \over {\partial t^2}} - 
  k_{n+1} ( y_{n+1} - y_n ) + 
  k_n ( y_n - y_{n-1} ) 
  \nonumber \\
  - 2  e^{-  y_n} 
  ( e^{-  y_n} - 1)  = 0 \; .
\end{eqnarray}
One notices that the last set of equations contain only one parameter,  the
coupling constant.  In order to represent the perturbation due to the enzyme,
one could imagine to modify locally any of the parameters of
Eq.~(\ref{discrete}), but it is likely that the presence of an enzyme will
affect the coupling constant through the bending of the molecule.  Moreover
previous studies of the role of disorder on the dynamics of the DNA
model\cite{Tashkent} have shown that the formation of open regions in the
model are much more sensitive to modulations of the coupling constant than
to changes in other parameters.  Therefore we only consider here an extended
perturbation of the coupling constant. An additional possibility to model
the enzyme specificity is however examined in the discussion.

Since we do not know how to solve in the discrete case, we transform the set
of equations, Eqs.~(\ref{ND}),  into the corresponding continuous PDE. In the
continuum limit, with a Taylor expansion in the potential term which assumes
small amplitude oscillation, Eq.~(\ref{ND}) becomes,
\begin{equation}
\label{continue}
   {{\partial^2 y} \over {\partial t^2}} 
  - {\partial \over {\partial x} } (k_1  
  {{\partial y} \over {\partial x}}  ) d^2
  + 2   ( y - 
  {3 \over 2} y^2 +
  {7 \over 6} y^3 ) = 0 \; ,
\end{equation}
in which $d$ is the lattice spacing and $k_1$ is a space dependent coupling
constant. We set $d$ equal  to unity in the following calculations.

\section{The Perturbed Nonlinear Schr\"odinger Equation}

Equation (\ref{continue}) can be transformed into a perturbed Nonlinear
Schr\"odinger equation by a multiple-scale expansion\cite{H,R}. Assuming
that the amplitude of the thermal oscillation is small, $y \approx
\epsilon \phi $, we perform the expansion
\begin{eqnarray}
\label{eq:approx}
\phi & = & F_0 + \epsilon F_1 + \epsilon^2 F_2 + O (\epsilon^3) \; , \\
{\partial \over {\partial t}} &  = &
{\partial \over {\partial t_0}} + {\partial \over {\partial t_1}} \epsilon +
{\partial \over {\partial t_2}} \epsilon^2 + O ( \epsilon ^3 ) \; , \\
{\partial \over {\partial x}} & = &
{\partial \over {\partial x_0}} + {\partial \over {\partial x_1}} \epsilon +
{\partial \over {\partial x_2}} \epsilon^2 + O ( \epsilon ^3 ) \; .
\end{eqnarray}
Moreover we assume a modulation of the coupling constant of the order of
$\epsilon$, {\it i.e.}
\begin{equation}
\label{assume}
{{\partial k_1}  \over {\partial x}} \approx
{{\partial k_1} \over {\partial x_1}} \; \epsilon \; .
\end{equation}
Equating like powers of $\epsilon$ yields a sequence of equations, in
ascending powers of $\epsilon$:
\begin{eqnarray}
\label{eq:order}
& & {\partial^2 F_0 \over \partial t_0^2} - 
k_0 {\partial^2 F_0 \over \partial x_0^2} + 2 F_0 = 0 \; ,\\
& & ({\partial^2 F_1 \over \partial t_0^2} +
2 {{\partial^2 F_0} \over {\partial t_0 \partial t_1}}) - 
{{\partial k_1} \over \partial x_1}{\partial F_0 \over \partial x_0} -
k_0 ({\partial^2 F_1 \over \partial x_0^2} +
2 {{\partial^2 F_0} \over {\partial x_0 \partial x_1}}) \nonumber \\
& &+ 2  (F_1 - {3 \over 2} F_0^2 ) = 0 \; ,\\
& &{\text{ and higher order equations}} \; , \nonumber
\end{eqnarray}
in which $k_0$ is the unperturbed coupling constant. Solving for equations
in each order of $\epsilon$ sequentially one obtains,
\begin{eqnarray}
F_0 &=& u ( x_1, x_2, t_1, t_2) e^{ i (q x_0 - \omega t_0 )} + c.c. \; ,\\
F_1 &=& {3 \over 2} | u |^2 + 
{{ 3  u^2 }\over { -4 \omega^2  + 4 k_1 q^2 + 2 }} 
e^{ 2 i (q x_0 - \omega t_0 )} \nonumber \\
& &+ c.c. \; ,
\end{eqnarray}
and the dispersion relation: $\omega^2 = \omega_0^2 + {k_0 } q^2 \;$, with
$\omega_0^2 = 2 $. From the vanishing of the secular equation at $ q = 0 $
one obtains the perturbed Nonlinear Schr\"odinger equation (NLS) at order
$\epsilon^2$:
\begin{equation}
 2 i \omega { \partial u \over \partial t_2} + 
{{\partial k_1} \over {\partial x_1}} 
{{\partial u} \over {\partial x_1}} +
k_1 {{\partial^2 u} \over {\partial x_1^2}} 
+ 8  u | u |^2 = 0 \;.
\end{equation}
We can further rescale the equation into a standard form:
Defining the following new dimensionless variables,
\begin{eqnarray}
\hat k ( \hat x ) &=& {k_1 \over k_0} - 1 \, ,\\
\hat u &=& \sqrt{ { 8 } \over { \omega^2}} u \, , \label{u}\\
\hat x &=& \sqrt{ { \omega^2} \over { 2 k_0 } }  x_1\, , \label{x}\\
\hat t &=& {\omega \over 2} t_2 \, ,
\end{eqnarray}
with $\hat k$ being the normalized deviation coupling in the vicinity of the
enzyme, one obtains the following perturbed dimensionless NLS,
\begin{equation}
\label{PNSE}
  i  {\hat u}_{\hat t}  + {1 \over 2} \hat u_{\hat x \hat x} + 
  \hat u | \hat u |^2 + {1 \over 2} 
{{\partial } \over {\partial {\hat x}}} ( {\hat k} 
{\hat u}_{\hat x} ) = 0 \;,
\end{equation}
and the corresponding Lagrangian density,
\begin{equation}
\label{Lagrangian}
\Lambda = {i \over 2} ( {\hat u}^* {\hat u}_{\hat t} - 
{\hat u} {\hat u}_{\hat t}^*) 
- {1 \over 2} ( { 1 }  + {\hat k} ) | {\hat u}_{\hat x} |^2
+ {1 \over 2} | {\hat u} | ^ 4 \; .
\end{equation}
In the following section we drop the $\;\hat{}\;$ for 
nomenclature simplicity.

\section{One soliton collective coordinate analysis.}

The collective coordinate method, which is a particle description of the
soliton in contrast to the field description given by the Lagrangian,
provides a good way to study the influence of a perturbation on a soliton.
The spirit is the same as  using the center of mass to analyze the behavior
of a system of particles. 

Without the perturbating term in Eq.~(\ref{PNSE}), one has a breather
solution
\begin{equation}
\label{breather}
u ( x , t ) = \eta
{\rm sech} 
 [  \eta ( x - u_e t ) ]
e^{i {u_e} ( x - u_c t )} + c.c.
\;,
\end{equation}
in which $\eta = \sqrt {(u_e^2 -2 u_e u_c)/(2PQ)}$,where $u_e$ is the
envelope  velocity, $u_c$ the carrier velocity, and $P = 1 / 2$, $Q = 1$
are coefficients of the second space  derivative and the nonlinear terms in
Eq.~(\ref{PNSE}) respectively. In view of this solution,  we use an {\em
ansatz} for the collective coordinate analysis
\begin{equation}
\label{ansatz}
u ( x , t ) = \eta\, {\rm sech} ( {{\eta x} } - \zeta )\, 
e^{i (\phi + \xi x )} \;,
\end{equation}
where the parameters $\eta$, $\zeta$, $\phi$, $\xi$
are functions of $t$. 
For an unperturbed system this implies the following relations between 
the parameters:
\begin{eqnarray}
\eta &=& \sqrt {{u_e^2 - 2 u_e u_c} } \, , \label{eta}\\
\zeta &=& {u_e \eta t }\, , \label{zeta}\\
\xi &=& {u_e }\, , \label{xi}\\
\phi &=& - {{u_e u_c t} } \,.
\end{eqnarray}
At $t = 0, \zeta = 0$ and $\phi = 0$ and there are only two parameters left,
which is consistent with Eq.~(\ref{breather}), because the NLS breather is a
two-parameter  solution. Even when the breather is far away from the defect,
because the ansatz extends to infinity and  always feels the defect, we do
not expect these relations to hold for a perturbed system. Hence in what
follows we examine the whole  four-parameter space for the equations of
motion.

Introducing this ansatz into the Lagrangian density,
Eq.~(\ref{Lagrangian}),  and integrating over space,  one obtains an
effective Lagrangian,
\begin{eqnarray}
\label{eqlagrange}
L &=& - 2 \eta \phi_t - 2  \zeta \xi_t + 
{{ \eta^3 } \over 3}  -  \xi^2 \eta 
 - {1 \over 2} \int_{- \infty}^{+ \infty}  k | u_x |^2 d x
\; .
\end{eqnarray}
and the corresponding Hamiltonian
\begin{eqnarray}
\label{Hamiltonian}
H &=& - {\eta^3 \over 3} + \xi^2 \eta 
 + {1 \over 2} \int_{- \infty}^{+ \infty}  k | u_x |^2 d x
\; ,
\end{eqnarray}
which contains no momentum term.

At this point, we must specify an expression for $k(x)$ to proceed. For 
algebraic convenience let us choose
\begin{equation}
\label{step}
 k = \kappa [ \Theta ( x     + l ) - \Theta ( x     - l ) ] \; ,
\end{equation}
in which $\Theta$ is the Heaviside step function and $l$ is the half-length
of the defect. This form of $k$ violates Eq.~(\ref{assume}),  however
previous works showed that  the collective coordinate results are generally
robust for the treatment of dynamics in the presence of
perturbation\cite{SB3}, therefore we can expect to get results which are at
least qualitatively correct in spite of this rather crude approximation.
Moreover we shall check them against full numerical simulations in the next
section.

Introducing the following abbreviated notation: 
\begin{eqnarray}
 T_+ &=& \tanh (\eta  l  + \zeta ) \, ,\\
 T_- &=& \tanh (\eta  l  - \zeta ) \, ,\\
 S_+ &=& {\rm sech} (\eta  l  + \zeta ) \, ,\\
 S_- &=& {\rm sech} (\eta  l  - \zeta ) \, , 
\end{eqnarray}
one obtains
\begin{equation}
 \int_{- \infty}^{+ \infty}  k | u_x |^2 d x = {\kappa \over 3} 
( T_+^3 + T_-^3 ) \eta^3 + \kappa ( T_+ + T_- ) \xi^2 \eta \; ,
\end{equation}
which characterizes the effect of the defect and decays fast towards zero
soon outside of the impurity region, and the equations of motion:
\begin{eqnarray}
\phi_t &=& {{ {\eta^2} \over 2} }  - { {\xi^2 \over 2} } 
 - { \kappa \over 4} ( T_+ + T_- ) \xi^2
- {{\kappa l } \over 4} ( S_+^2 + S_-^2 ) \xi^2 \eta
\nonumber \\
& &- {{\kappa l } \over 4} ( S_+^2T_+^2 + S_-^2T_-^2 ) \eta^3
- {{ \kappa} \over 4} ( T_+^3 + T_-^3 ) \eta^2 \, ,
\\
\xi_t &=& 
- {{\kappa } \over 4} ( S_+^2T_+^2 - S_-^2T_-^2 ) \eta^3
- {{\kappa } \over 4} ( S_+^2 - S_-^2 ) \xi^2 \eta \, , \\
\zeta_t &=&  \xi \eta + {{\kappa} \over 2} ( T_+ + T_- ) \xi \eta \, ,
\label{CZ}\\
\eta_t &=& 0 \label{CA} \, . 
\end{eqnarray}

As expected, far away from the defect, {\it i.e.} when $S$ and $T$ vanish,
one recovers the usual relations for the NLS equation because in this case
$\xi_t = 0$ so that $\xi$ is a constant that we can denote by $u_e$.  Then
$\zeta_t =  \xi \eta$ gives $\zeta =  u_e \eta t$ as expected,  and $\phi_t
=  ( \eta^2 -  \xi^2) / 2$ gives 
$\phi =  -  u_e u_c t $ if $u_c$ is defined by Eq.~(\ref{eta}) for $\eta$.

In the presence of the defect, the set of nonlinear differential equations
for the collective variables cannot be integrated analytically. It is
however much simpler than the full set of discrete equations since it
contains only 3 equations. It can be integrated by a  fourth-order
Runge-Kutta method. One can however make general remarks on the properties
of the solution before resorting to numerical calculations. The soliton
described by the ansatz is an unbreakable entity and moreover  the energy
given by Eq.~(\ref{Hamiltonian}) is conserved even when a potential well is
encountered.
Therefore when the soliton reaches a defect, it may speed up to compensate
for the extra energy requirement due to a decrease in coupling energy, as
shown in Fig.~\ref{CC}(a), in which a large carrier velocity was chosen to
exaggerate the effect. However, the behavior is richer than the one
generally found for topological solitons because, in addition to the time
dependent position $\zeta(t)$, the ansatz contains an internal degree of
freedom, $\xi$, so that the energy can also be transferred between different
collective coordinates. With $\kappa < 0$ the last term of
Eq.~(\ref{Hamiltonian}) decreases in the region of the defect,  therefore
$\xi^2 $ has to increase accordingly.  If $\kappa > 0$ and the breather is
initially inside  the defect, it simply slips away as in Fig.~\ref{CC}(b). 
If it is initially outside,  for some suitable range of amplitude it is
first slowed down and eventually reflected, as shown in Fig.~\ref{CC}(c);
reflection occurs beyond $\eta \approx 0.25$. For smaller $\eta$, breathers
pass through the defect, indicating that the broader breathers are less
influenced by the presence of defects, just as a large-wheel bike will not
be stopped by a pebble or a ditch.  In Fig.\ref{CC}(c) the breather has
actually penetrated into the defect before being reflected.  When the
breather is trapped it oscillates between two positions, which may not be
the defect boundary, as shown in Fig.~\ref{CC}(d). For values of $\eta$ close
to the threshold between trapping and non-trapping, the breather slowly
turns around at the boundaries as in Fig.~\ref{CC}(e).


A {\it necessary} condition for a moving breather to be trapped in the above
defect is $\kappa < 0$. This statement can be proved through the following
argument: a  necessary condition for trapping is $\zeta_t = 0$ more than
twice,  which, according to Eq.~(\ref{CZ}), is equivalent to 
\begin{equation}
\cosh^2 \zeta = 
1 - \cosh^2 ( \eta l ) - {\kappa} \sinh ( \eta l ) \cosh ( \eta l ).
\label{trap}
\end{equation}
Since $\cosh^2 > 1$, $k_0 > 0$ and $\eta > 0$, 
$\kappa $ has to be less than zero.
We have therefore showed that trapping occurs only if the  perturbed
coupling constant  is less than the unperturbed one,   which is consistent
with our simulations although it has been proven only in the collective
coordinate approach.

Since Eq.~(\ref{trap}) contains only $\zeta$, $\eta$, $\kappa$, and $l$, if
the characters of the defect,  {\it i.e.} the length, $l$, and the strength,
$\kappa$,   have been fixed for a given system, and the initial position of
the breather is chosen, the only factor which characterizes trapping is the
breather amplitude.  The initial value of $\phi $ seems to have no
consequence on the results. In general, if one finds that a breather passes
through a defect  for
$\kappa < 0$ as in Fig.\ref{CC}(a), one can obtain trapping by increasing
its amplitude.

Because of the helicoidal structure of DNA, a given strand is alternatively
inside and outside the bend so that it experiences a periodical modulation
of its elasticity by an attached enzyme. We examined the consequence of such
a modification by considering the following coupling constant modulation:
\begin{equation}
k = \kappa [ \Theta ( x + l ) - 2 \Theta ( x ) + \Theta ( x - l ) ]
\label{SS}
\end{equation}
for which the coupling is first increased by $\kappa$ and then decreased by
the same amount if $\kappa > 0$.  It can be viewed as a step approximation
of one period of a sinusoidal modulation. In general one finds nothing
essentially new  with this perturbation because it is only a superposition
of two step defects. However, this perturbation is asymmetric in space.  By
changing the sign of $\kappa$ we can reverse the orientation of the
perturbation with respect to an incoming breather. If $\kappa < 0$, a breather
starting from the left side of the defect  encounters first the region where
the coupling constant  is decreased. 
Table I summarizes the behavior of breathers with various amplitudes  and
the two possible signs of $\kappa$ and $\xi$.  For negative $\kappa$ and
positive $\xi$, the breather is reflected for intermediate $\eta$ while for
large enough $\eta$ it is trapped. If one switches to positive $\kappa$
there is still a range of $\eta$ values that produce reflection, but for
large $\eta$ the breather passes through. If the breather starts from the
side where coupling constant is decreased the trapping can still exist even
if the initial position of the breather is far away from the defect, but the
pass-through region disappears as expected. These results show that it is
the first encounter which determines the trapping. However, for this case of
a composite defect the collective coordinate calculation can, in some cases,
lead to qualitatively wrong results. The full numerical calculation shown in
Fig.\ref{all}(a) indicate that the breather can be trapped even if it were
coming from the higher side of the defect. This points out the limit of the
collective coordinate method for successive perturbations of the breather.
The first interaction of the breather with a perturbation appears to be
qualitatively well described. But then the perturbed breather is not
accurately described by the ansatz. Thus when it encounters a second
perturbation (here the second step in coupling constant), the collective
coordinate description fails to describe the interaction.

\section{Direct numerical simulations}

Since the last example has shown that the collective coordinates  cannot
provide a full description of the breather dynamics, it is necessary to
check them against full numerical simulations of Eqs.~({\ref{ND}). Using the
breather solution given by Eq.~(\ref{breather}) as an initial condition,
and  periodic boundary conditions, we integrate Eqs.~({\ref{ND}) with a
4$^{\text{th}}$ order  Runge-Kutta scheme and a time step chosen to provide
a conservation of energy to an accuracy better than $10^{-6}$ over a full
simulation. The calculations have been tested on different system sizes to
make sure that the results are not modified by boundary effects. The ansatz
(\ref{breather}) is not an exact solution of the full set of equations
because the transformation to the NLS form involved several approximations,
however, except for very discrete cases or large amplitude breathers, it
provides a rather good solution far away from the defect. As long as the
breather is far away from the defect, one generally notices only a small
decay of  the initial energy peak due to radiation. 

Full numerical calculations  have the advantage of allowing radiation and
breaking of a breather. Furthermore, although the collective coordinate
method starts from the perturbed NLS which requires small $u_e$  and $u_c$
and hence small amplitude, the full numerical calculations  do not have this
restriction. In what follows we show both energy distribution and the
breather amplitude. The energy distribution is more relevant to the  opening
of DNA chain while the breather amplitude allows a comparison with the
results of the collective coordinate calculations.

Fig.~\ref{DS}(a) is a typical case for trapping  at an equivalent amplitude
$\eta = 0.19$.  The correspondence is made from Eq.~(\ref{eta}) and
Eq.~(\ref{xi}). The threshold for trapping predicted by collective
coordinate is higher ($\eta = 1.01$  for a breather  initially at
$x = 12$ when both systems have the same  dimensionless coupling constant).
As noticed earlier, it is not surprising to find such a discrepancy because
we have used a sharp perturbation that violates the condition (\ref{assume}).
However the full simulations confirm the qualitative predictions of the
collective coordinate calculations: small amplitude breathers are
transmitted, while larger ones are trapped. Fig.~\ref{DS}(b) shows that the
energy distribution around the breather gets  sharper in the region of
the defect. Therefore a negative perturbation, which tends to trap
breathers, is also favorable for base-pair opening since it concentrates the
energy of the incoming breathers in a narrow domain. This sharpening of the
breather shape occurs when the breather is inside the perturbation domain,
whether it will stay trapped or not. One can also find  on the contrary
that, if a breather  meets a positive perturbation, its energy distribution
broadens. This behavior is similar to that of a vortex in shallow water: the
vortex becomes wider  when it is in shallower water and thinner in deeper
water.\cite{T} In the amplitude plot of Fig.~\ref{DS}(b),  when the breather
reaches the boundary of the defect, one can see two small reflected waves.
They were not included in the collective coordinate analysis, and their
presence explains part of the quantitative discrepancy between the
analytical approach and the full simulations. Sometimes one can also notice
that the breather changes its oscillation frequency  after the collision
with the defect. 

The results of the full numerical simulations show that, although the
collective coordinate analysis is able to predict qualitatively the main
features, in particular the existence of a threshold for trapping when the
breather amplitudes increases, it is quantitatively wrong. The same
conclusion had been found for an isolated impurity\cite{FPM}. There are
several reasons for that. Firstly we do not know an appropriate ansatz for
the original equations of motion (\ref{discrete}) and we start from a
perturbed NLS Lagrangian which is already approximate. Then we use an ansatz
which is localized in space and does not allow for the breaking of the
breather or the emission of reflected waves.  And finally the calculation
assumes a smooth evolution of the coupling constant while we later use a
sharp variation to make the analytical calculation possible.  In spite of
all their weaknesses, the collective coordinate calculations are however
useful to get an insight of the behavior of the breather in the presence of
the defect or even draw general conclusions on the kind of defects that can
trap energy as explained above.

Another point of interest is the trapping of {\it several} breathers in the
defect region which could really enhance the energy density locally and
cause local openings in DNA.  We show in Fig.~\ref{all} examples for trapping
for two kinds of coupling constant shapes:  Fig.~\ref{all}(a) is an example
when the breather comes from the higher side of a two-step defect but is
trapped. However, unless in favorable conditions we seldom find that two
breathers can be trapped inside the same perturbation.
When the second breather gets trapped it often kicks out the first breather
that was trapped before, as shown in Fig.~\ref{all}(b). In other cases we
noticed that, when a first breather is trapped in the defect, a second
breather that would have been trapped if it were alone, is on the contrary
reflected. Therefore, if one studies only the positions of the breathers
during their first interactions with the extended defect, it seems that the
defect will never collect more than the energy of one breather. This is in
fact not true, but the complete phenomena require a more detailed analysis.
It is interesting to study the evolution of the energy in the
region of the defect versus time. 
An example is shown on Fig.~\ref{figu5}. In this
case the first breather that interacts with the defect has an amplitude
$\eta = 0.2$  which is above the trapping threshold and the second one has an
amplitude $\eta = 0.1$ below the threshold. As expected the first breather
is trapped and oscillates around the defect. The second one passes
through the defect region that contains the first breather. However, if one
looks at the energy density on the 
three-dimensional plot of Fig.~\ref{figu5}(a), one
can notice a significant increase of energy density after the interaction of
the second breather with the defect. The reason is that the second breather
is only {\it partly} transmitted. A large part of its energy is given to the
trapped breather, i.e. it stays in the defect region. The same phenomenon
occurs again when the second breather collides a second time with the
trapped breather. Due to this complex process, the time evolution of the
energy inside the defect region 
(Fig.~\ref{figu5}-c) is a complicated curve, but it is
important to notice that it tends to grow, and never falls again to a small
value, indicating that the multiple collision process does cause a
concentration of energy in the defect region. The origin of this
localization of energy does not lie in breather trapping but in breather
interactions in the presence of a perturbation, and therefore it is not
included the collective coordinate description of Sect.~IV. The result is
very reminiscent of a mechanism described recently for energy localization
due to discreteness effects in nonlinear lattices\cite{Nlenloc}. In both
cases the collisions of breathers, perturbed either by a defect or by
discreteness, cause energy transfers that, on average, favor the big
excitation at the expense of the small one. We have checked that the
mechanism  is not restricted to a particular case. 
Fig.~\ref{figu6} shows another
example in which 3 breathers with the same initial amplitude
$\eta=0.2$ were sent to the defect. Although the details of the process are
different, they lead to the same final result: breather interactions in the
presence of the defect tend to favor the formation of a large amplitude
breather that concentrates a large part of the energy of the three incoming
breathers and is finally trapped at the defect site. Hence the energy in the
defect region settles to a high value. Tests have been performed with
various breather amplitudes, leading to the same general result.

\section{Conclusion}

Using a simple DNA model we have modeled the effect of a transcription enzyme
by an extended modification of the coupling constant along the strands. The
results show that such a perturbation is more efficient than an isolated
impurity to trap breathers, in particular because trapping can occur provided
that the amplitude of the incoming breather exceeds a threshold instead of
requiring breathers with a well defined frequency. This conclusion can be
derived from collective coordinate calculations as well as from numerical
integration of the full set of equations of motion, although the collective
coordinate method overestimates the trapping threshold. One cannot expect
quantitative results from the collective coordinate analysis because we have
violated at least one basic assumption, Eq.~(\ref{assume}), to allow the
analytical calculations, but it gives insight into the physics, and in
particular a necessary condition for breather trapping which is confirmed by
the full simulations.

We have also showed that energy exchanges between a first breather, already
trapped, and other incoming breathers can lead to a concentration of energy
in the region of the defect.

One may wonder whether the results obtained above for specific perturbations
are extendable to more realistic cases.  Although it is difficult to give
general answers to this question,  one can get insights through numerical
simulations of the full system. In real DNA one has 
$D = 0.03 eV$,  $\alpha = 4.45 \AA^{-1}$, $k_1 = 0.08 eV/\AA^2$, 
$m=300$ a.m.u. for AT base pairs, while for GC pairs we have $D=0.035 eV$, 
$k_1 =0.104$. \cite{Dauxois} This is equivalent to $k_n \approx 0.13$ to
$0.15$. In this range of coupling large amplitude breathers are trapped by
discreteness\cite{BangP}. The collective coordinate calculations suggest
that the low amplitude breathers, which can move, will not be trapped by the
20 base-pair defect. Simulations show that it is not necessarily so. For
instance a 20 base-pair defect with $K_n=0.12$ in a chain with $K_n=0.15$ 
can trap breathers of various amplitudes. The energy exchange mechanism in
the presence of the defect, discussed above, interferes with discreteness
effects that can have similar effects to localize energy\cite{Nlenloc}.
Therefore, although we have exhibited a mechanism which is active in a wider 
frequency range than an isolated defect, the calculations performed on a
simple model are not be sufficient to draw a conclusion about its validity
to describe the effect of an enzyme on DNA transcription. It may however
deserve attention because of its greater efficiency compared to the case
of a point defect that was considered previously. 

In this work we have modeled the role of the enzyme by modulating only the
coupling constant along the strands. As mentioned in the introduction, other
possibilities could be considered, particularly if one attempts to take into
account the  enzyme specificity which suggests that the enzyme could have
another role than merely bending locally the molecule. As a first step in
this direction, we have considered a local change of the Morse potential in
addition to the effect of the bending. 
Figure \ref{figu7} shows the result of a
numerical simulation where all the conditions are the same as for
Fig.~\ref{figu6},
except that, in addition to changing the coupling constant inside the defect
to model the bending, we have also multiplied the denaturation energy of the
base pairs (parameter $D$ of Eq.~(\ref{discrete})~) by a factor 0.8. This
means that we also assume that the enzyme can have some chemical effect to
reduce the base pairing interaction. The comparison of Fig.~7 and Fig.~6c
shows that this modification has a rather drastic effect on the results. It
is easy to understand qualitatively why because locally the vibrating
frequency of base pairs has been reduced. As we consider low energy
breathers which are the most likely to be excited at physiological
temperatures, their frequency, situated below the base-pair linear frequency
of the unperturbed region because of the soft nonlinearity of the Morse
potential, is however very close to the bottom of the phonon band of the
unperturbed part of the molecule. As the enzyme lowers the frequencies of
phonon band in the defect region, the breather frequency is now {\it in
resonance} with some modes of the phonon band of the defect. Therefore when
the breather is trapped at the defect site by the bending, it is trapped in
a region where it resonates with phonons. As a result it loses energy by
radiation, but, as the emitted modes have a frequency below the lowest
frequency of the unperturbed lattice, the vibrations are trapped in the
defect region. One observes that the trapped breather spreads out its energy
inside the defect region. When a second breather comes to this excited
defect it is no longer repelled by a highly 
localized breather as in Fig.~\ref{figu6}.
Thus it is more likely to penetrate in the defect region too. This makes the
energy localization effect more efficient and instead of the large
oscillations that were observed in Fig.~\ref{figu6}-c, 
Fig.~\ref{figu7} shows that the energy in
the defect region now grows steadily, each new breather having a high
probability to add its contribution. Although it is still preliminary, this
example shows that, if one combines the bending effect of the enzyme with
some model for its specific action on the promoter site, one can perhaps
provide a mechanism to achieve the local opening of the double helix which
is required by DNA transcription.

\section{Acknowledgments}
We would like to thank T. Dauxois for helps on computer programming.
J.J.-L.T. acknowledges the hospitality of the Laboratoire de
Physique de l'Ecole Normale Sup\'erieure de Lyon where part of this work
was done. J.J.-L.T. also acknowledges 
partial support of National Science Council, Taiwan, 
grant No. 84-2911-I-007-030-B21. Part of this work has been supported by the
EU Science Program through grant SC1*CT91-0705.


\begin{figure}[bth]
\begin{center}
\mbox{\psboxto(7cm;3cm){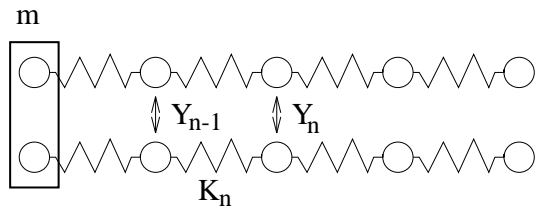}} \\
\caption{A one-dimensional lattice model of DNA.}
\label{line1}
\end{center}
\end{figure}

\begin{figure}[bth]
\begin{center}
\mbox{\psboxto(6.cm;7.5cm){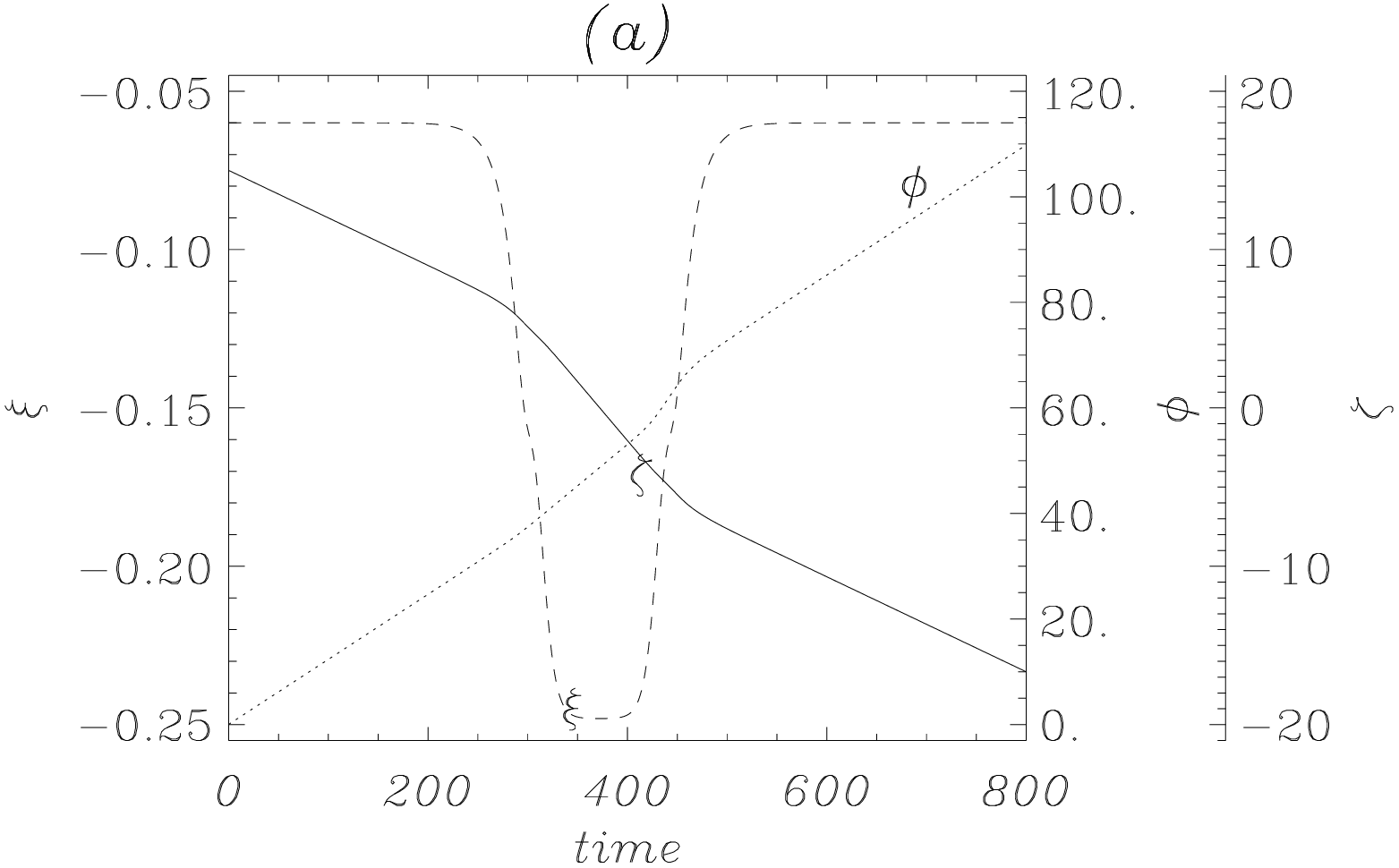}} \\
\mbox{\psboxto(6.cm;7.5cm){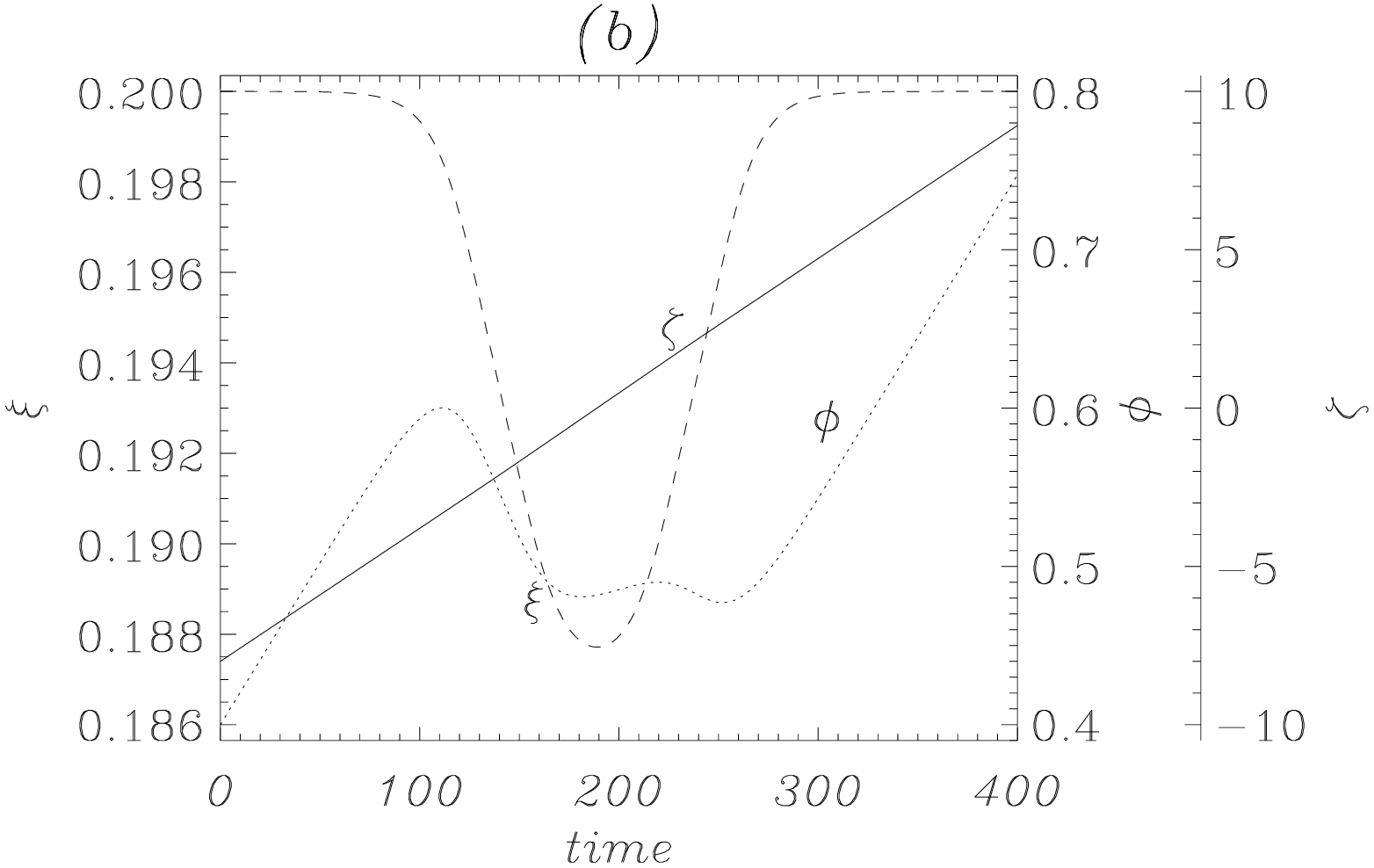}} \\
\mbox{\psboxto(6.cm;7.5cm){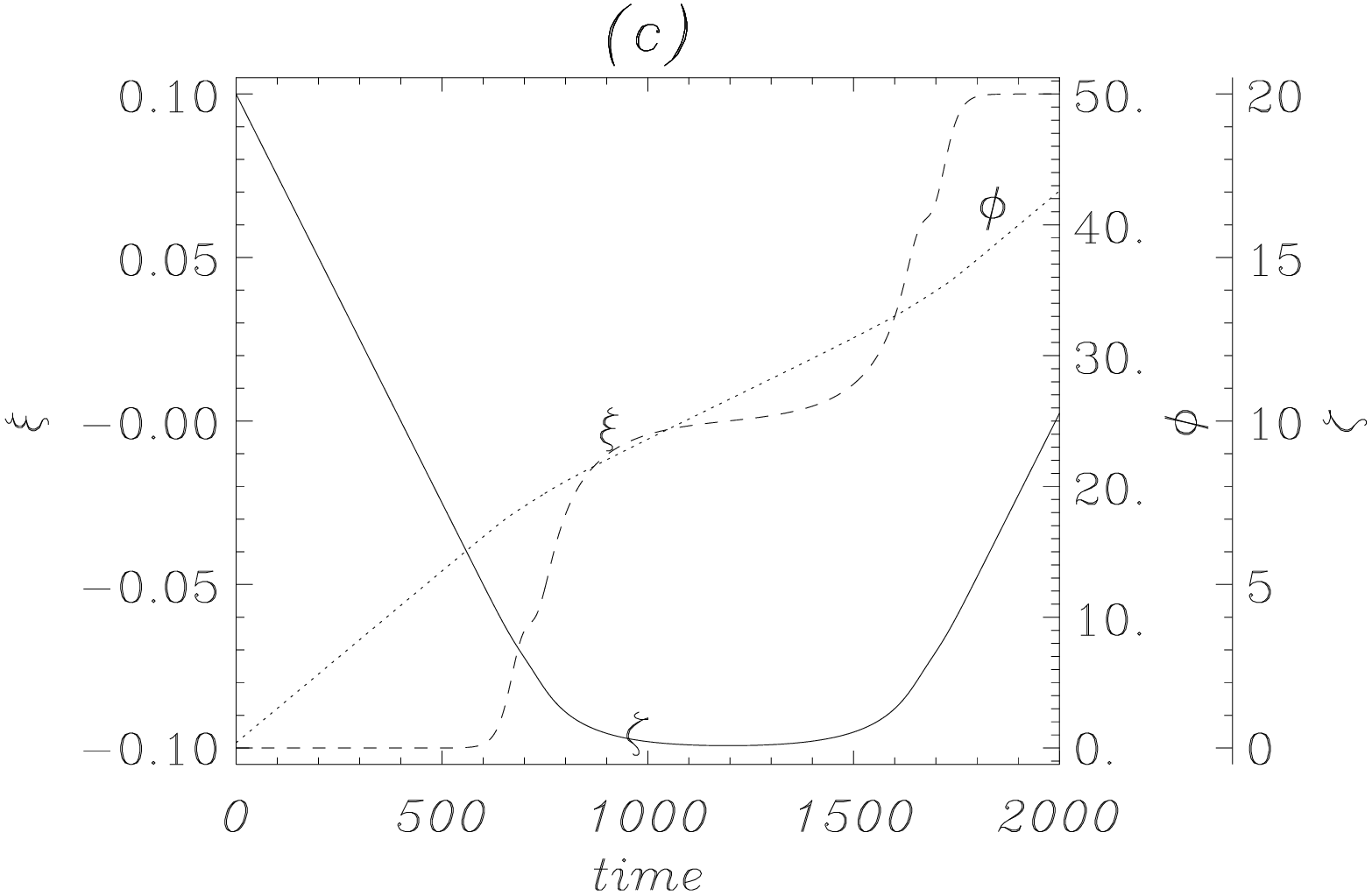}} \\
\mbox{\psboxto(6.cm;7.5cm){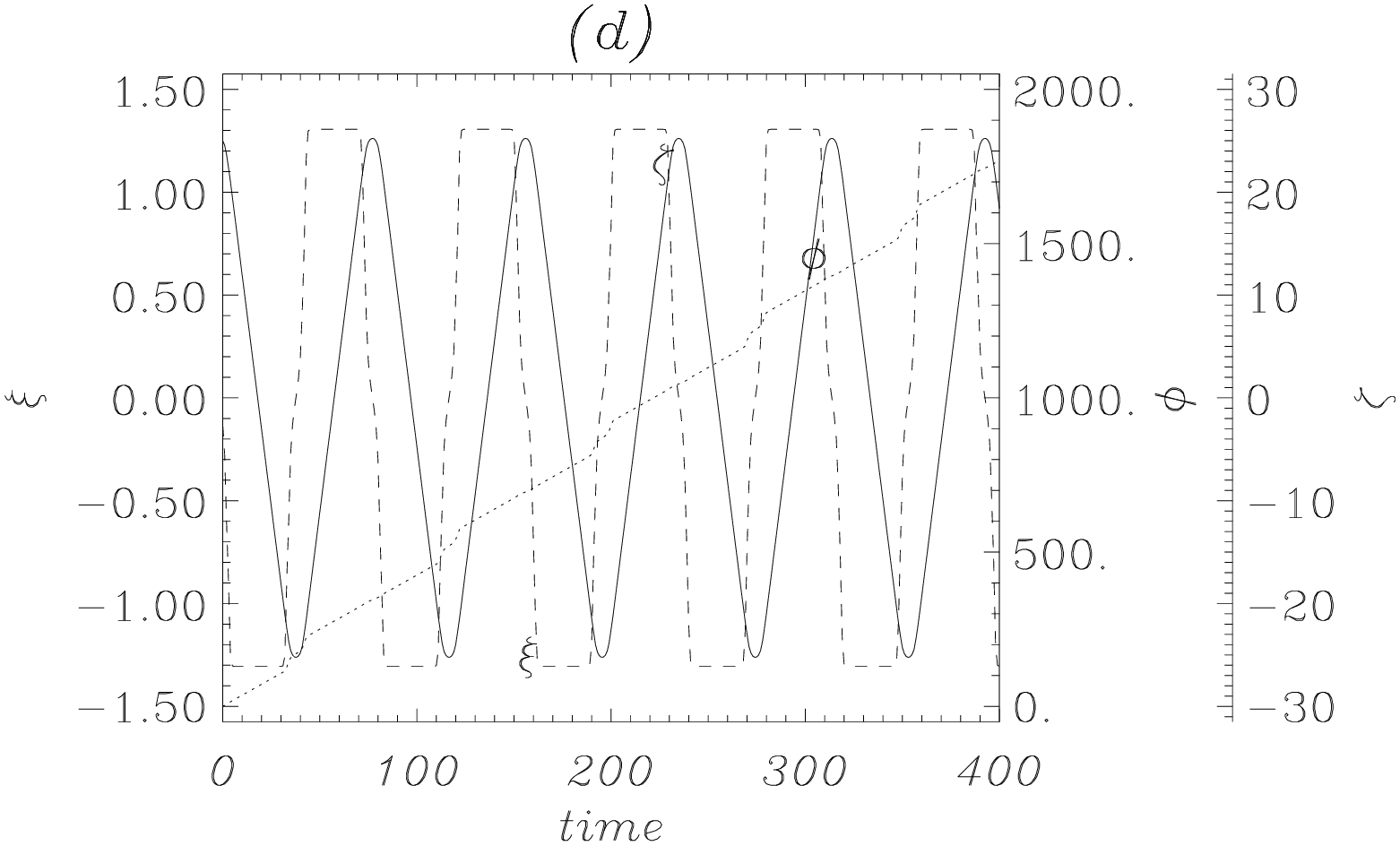}} \\
\mbox{\psboxto(6.cm;7.5cm){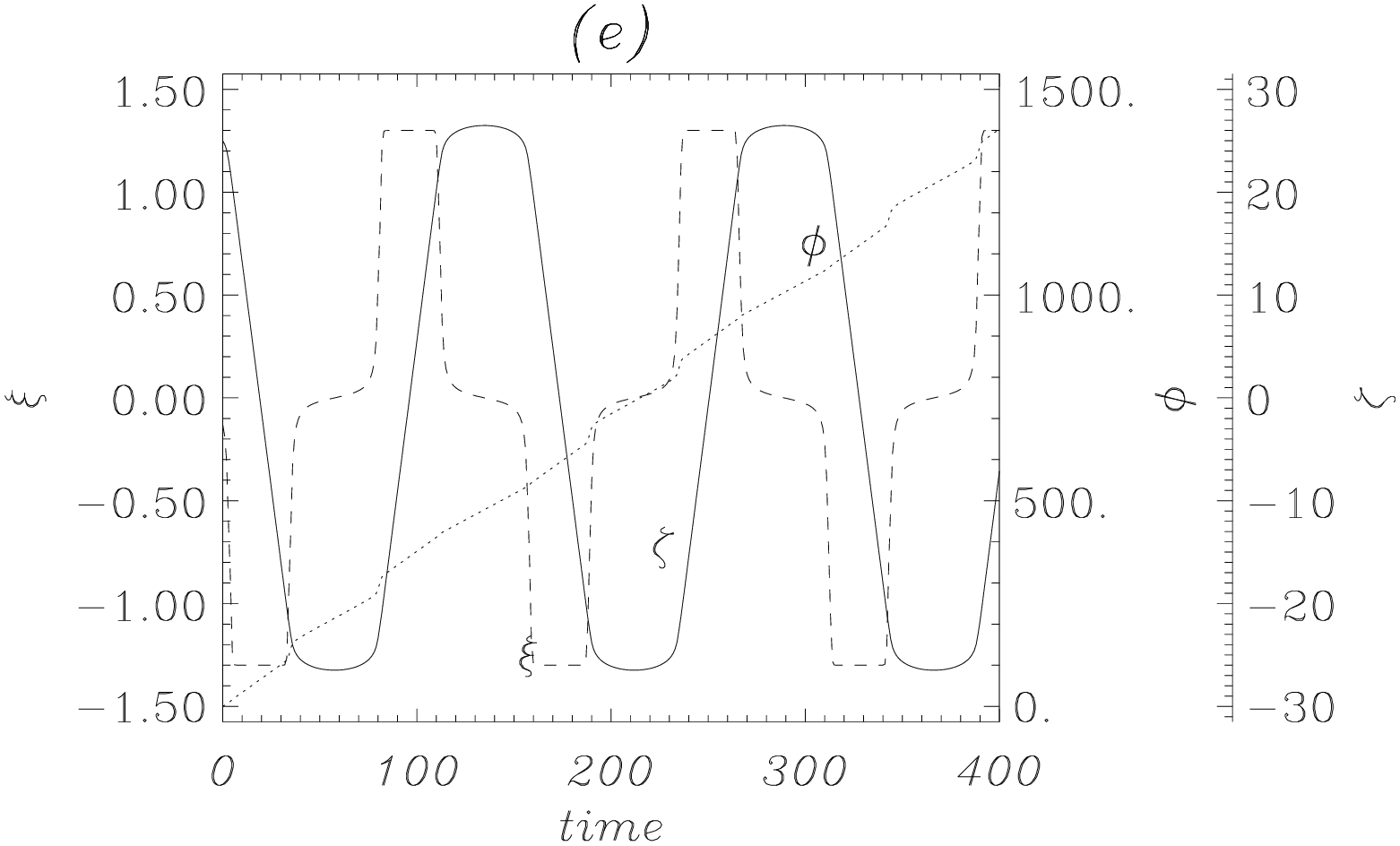}} \\
\end{center}
\caption{Evolution of collective coordinates in the presence of a step
defect, Eq.~(\protect\ref{step})
for $l = 10 $, 
(a) $\eta = 0.5$, $ \kappa = -0.4$, $\phi ( 0 ) = 0.05$, $\xi ( 0 ) = -0.06$, 
$ \zeta ( 0 ) = 15$. 
(b) $\eta = 0.21$, $ \kappa = 0.1$, $\phi ( 0 ) = 0.4$, $\xi ( 0 ) = 0.2$, 
$ \zeta ( 0 ) = -8$. 
(c) $\eta = 0.25 $, $ \kappa = 0.5$, $\phi ( 0 ) = 0.4$, $\xi ( 0 ) = -0.1$, 
$ \zeta ( 0 ) = 20$. 
(d) $\eta = 2.3$, $ \kappa = -0.5$, $\phi ( 0 ) = 0.1$, $\xi ( 0 ) = -0.13$, 
$ \zeta ( 0 ) = 25$. 
(e) $\eta = 2.253$, $ \kappa = -0.5$, $\phi ( 0 ) = 0.1$, $\xi ( 0 ) = -0.13$, 
$ \zeta ( 0 ) = 25$.  }
\label{CC}
\end{figure}

\begin{figure}[bth]
\begin{center}
\mbox{\psboxto(19cm;21cm){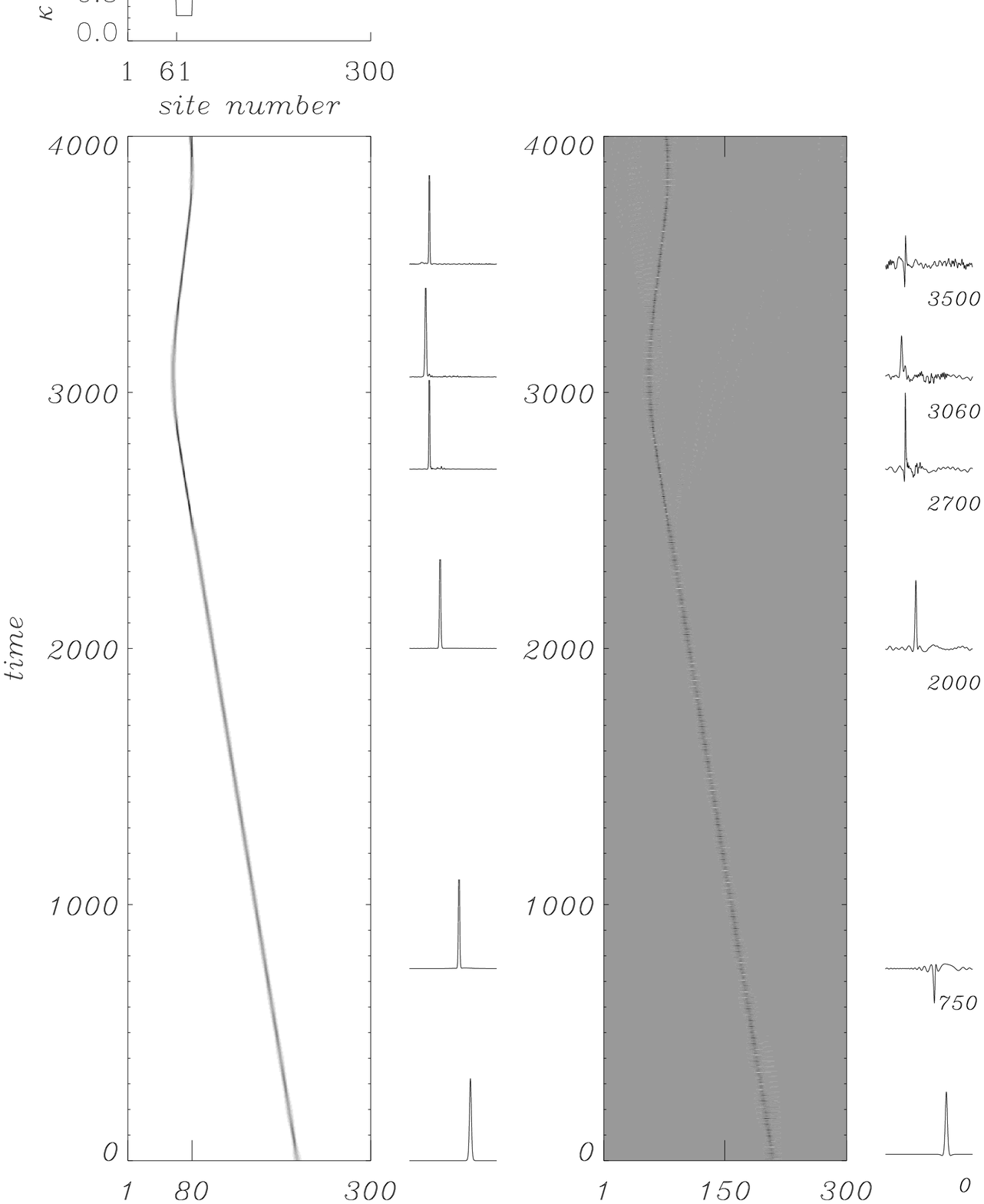}} \\
\mbox{\psboxto(19cm;21cm){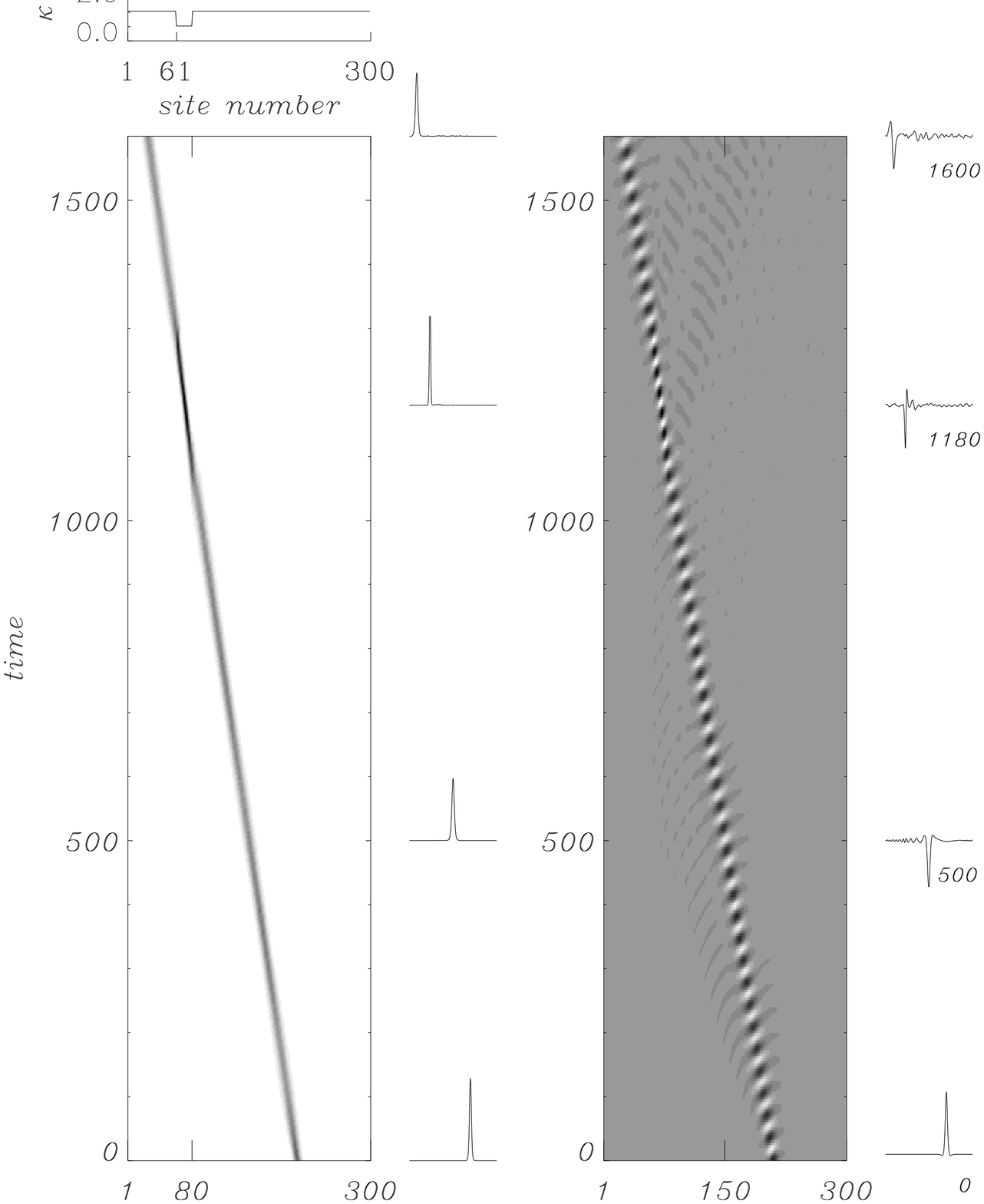}} \\
\end{center}
\caption{Half-tone plots of the energy distribution of 
breather evolutions, left, and the corresponding amplitudes, right, from 
direct numerical integration of Eq.~(\protect\ref{ND}).
In each figure the top insert
shows the variation of the coupling constant used for the calculation, while
the right inserts show snap-shots of the breather 
energy or amplitude distributions.
Defects positions are shown in the plots axis. 
(a) $K_n = 0.44$ outside the defect, $ K_n = 0.22$ for the defect. 
$ X ( 0 ) = 210$, $u_c = 0.13$, $u_e = -0.1$. 
(b) $K_n = 1.04$ outside the defect, $ K_n = 0.52$ for the defect. 
$ X ( 0 ) = 210$, $u_c = 0.13$, $u_e = -0.1$. }
\label{DS}
\end{figure}

\begin{figure}[bth]
\begin{center}
\mbox{\psboxto(19.cm;21.cm){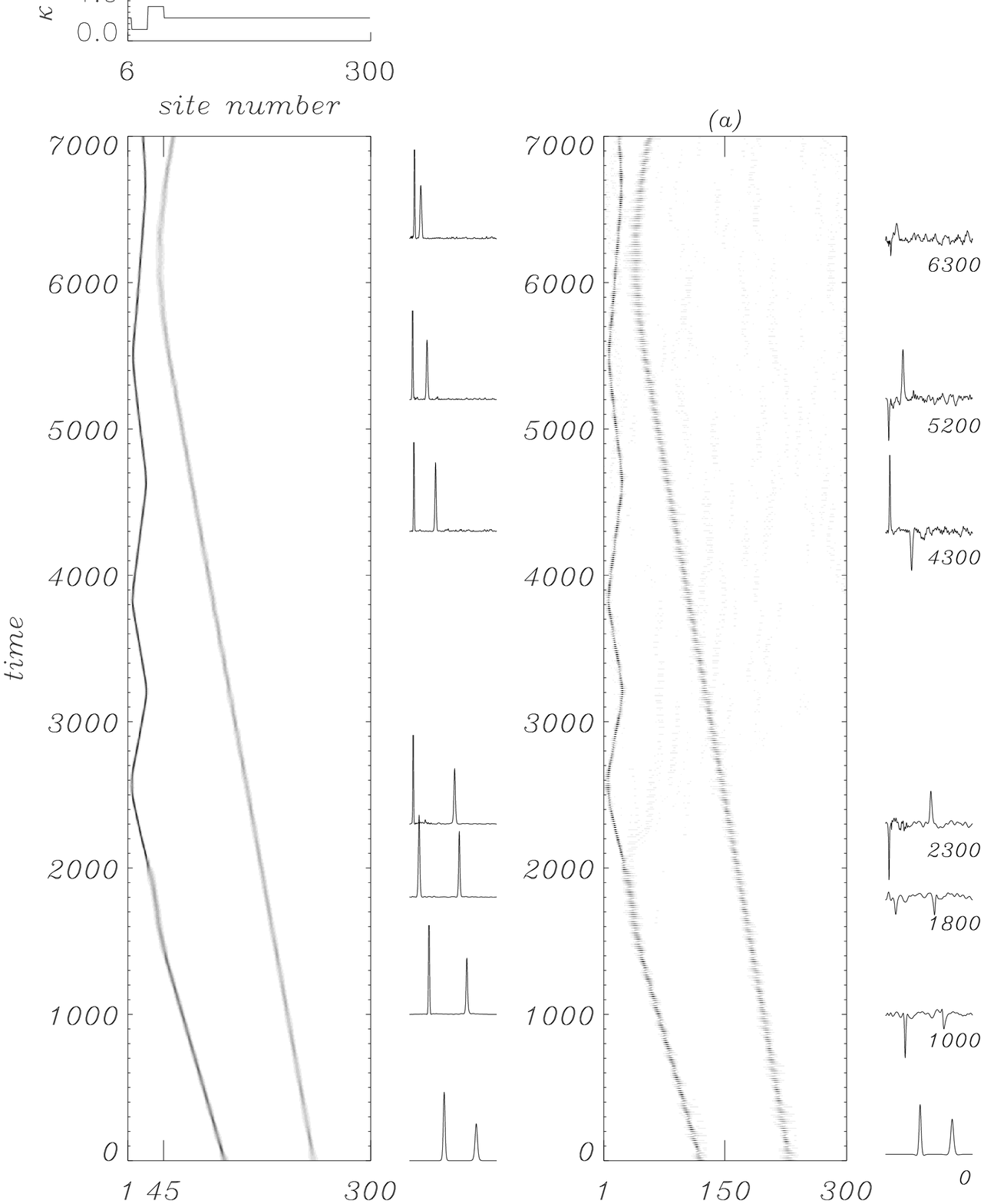}} \\
\mbox{\psboxto(19.cm;21.cm){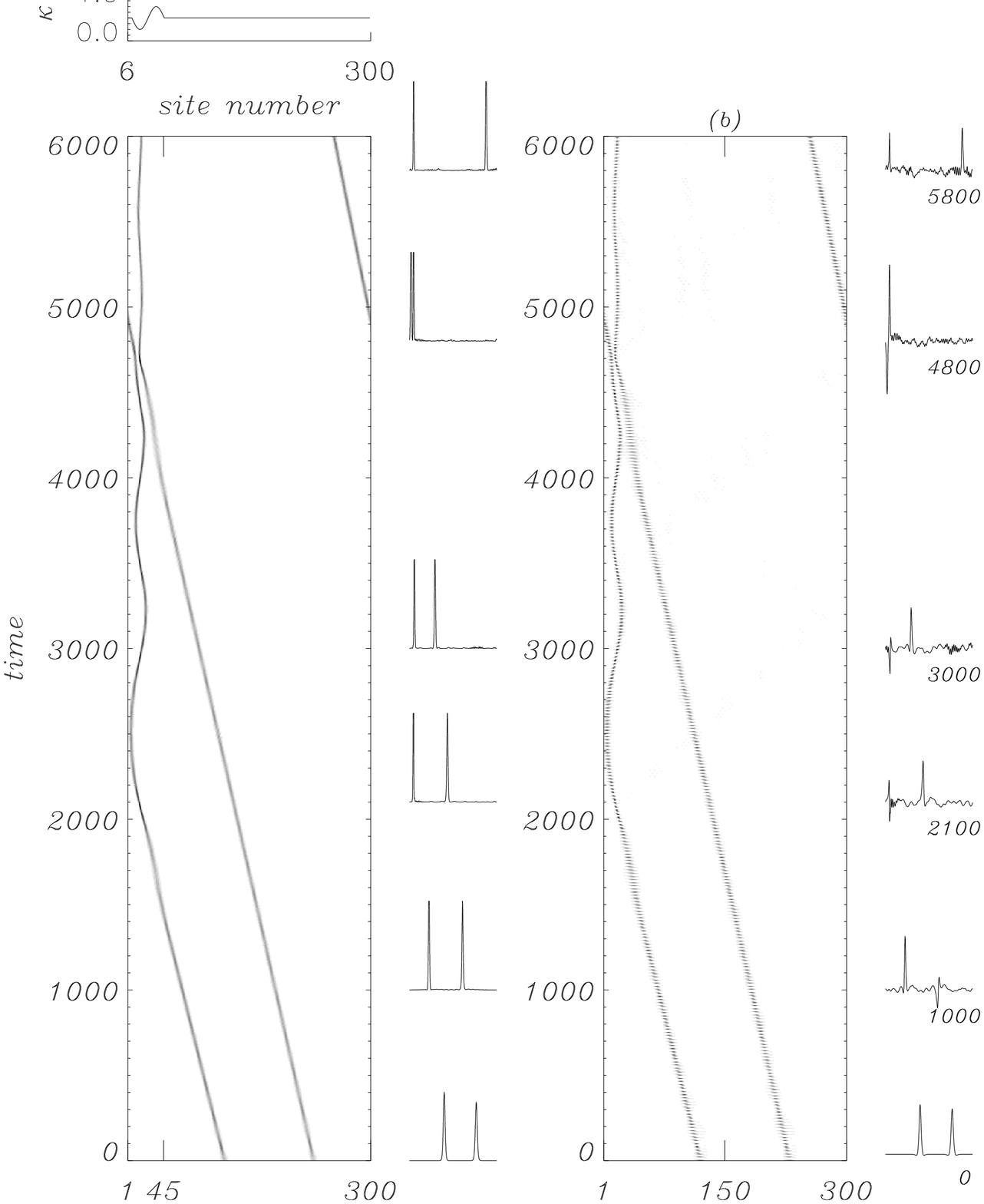}} \\
\end{center}
\caption{Half-tone plots of the energy distribution of two-breather 
evolutions, left, and the corresponding amplitudes, right, from 
direct numerical integration of Eq.~(5) 
and various  shapes of defects.
In each figure the top insert
shows the variation of coupling constant used for the calculation, while
the inserts show snap-shots of the breather 
energy or amplitude distributions.
The breathers start from $X ( 0 ) = 120, 230$,
defects positions are shown in the plots axis and 
(a) $K_n = 0.4$ outside the defect, $ K_n = 0.2 $ inside the defect and
$u_c = 0.13, 0.11$, $u_e = -0.11, -0.07$. 
(b) $K_n = 0.4$ outside the defect, $ K_n = 0.2 \sin ( n \pi / l) $ 
for the defect with
$n$ the distance from the center of the defect and
$l$ the half-length of the defect,  and
$u_c = 0.13, 0.12$, $u_e = -0.11, -0.1$. 
}
\label{all}
\end{figure}

\begin{figure}[bth]
\begin{center}
\mbox{\psboxto(7.cm;7.cm){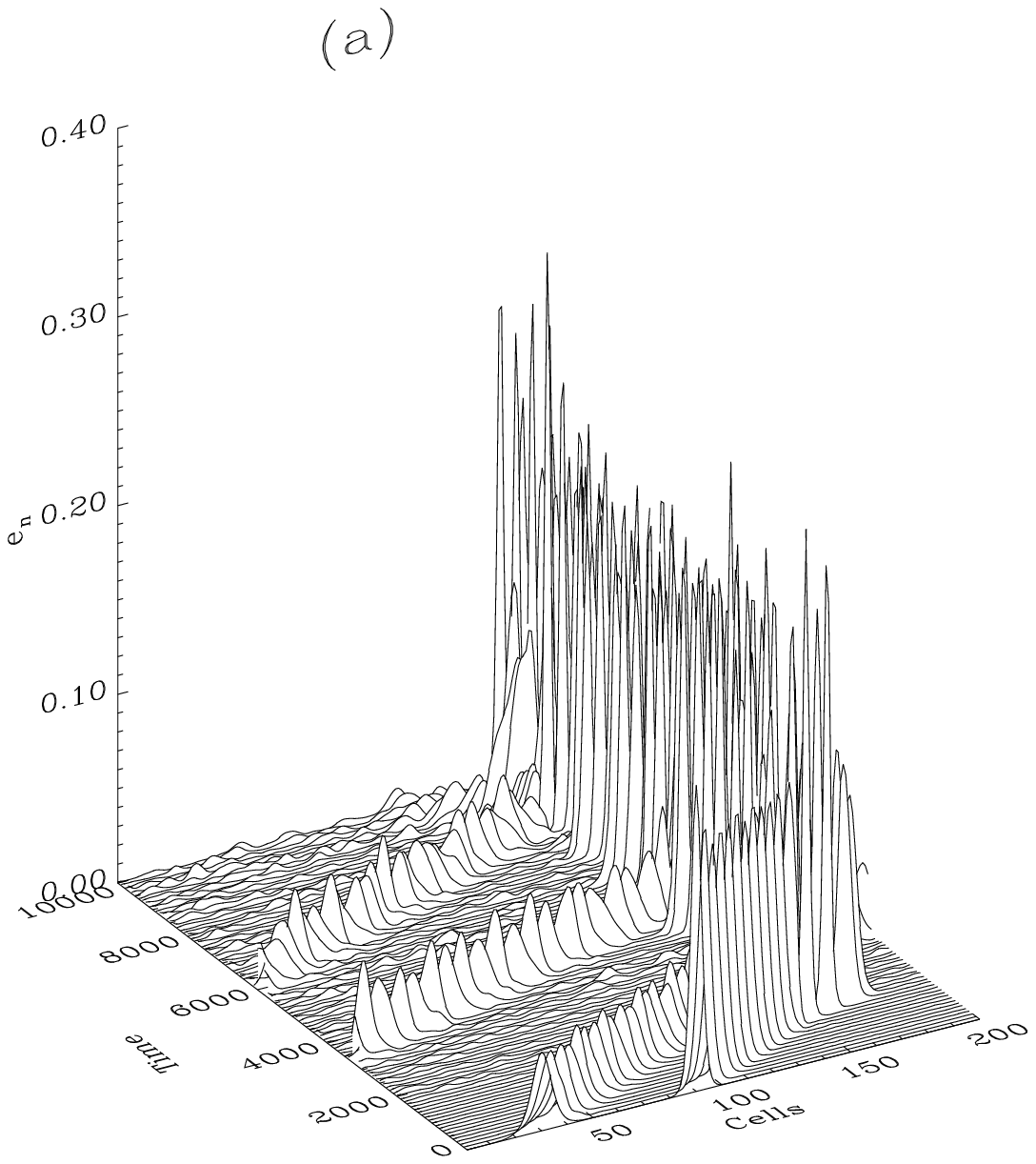}} \\
\mbox{\psboxto(7.cm;7.cm){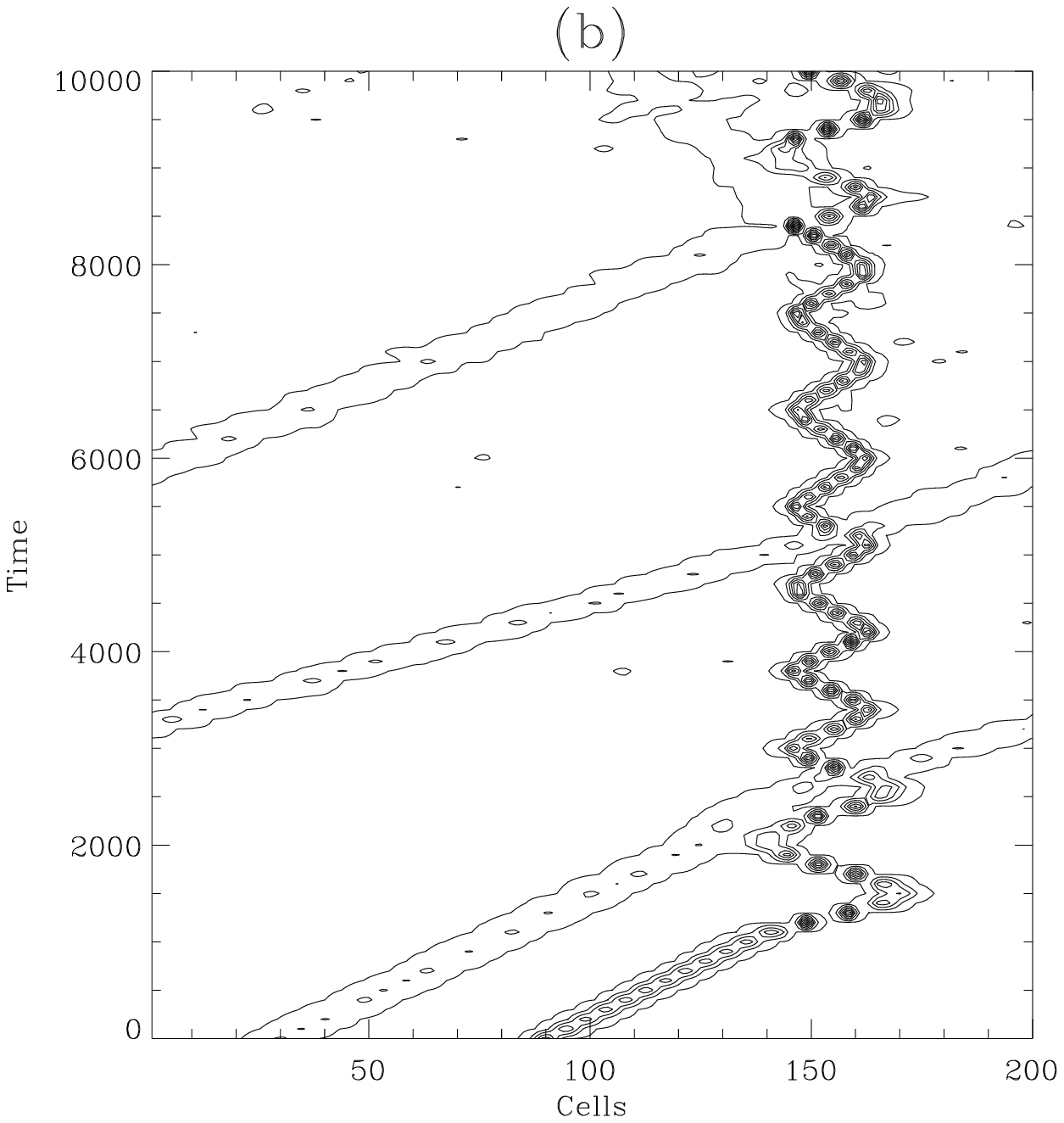}} \\
\mbox{\psboxto(7.cm;7.cm){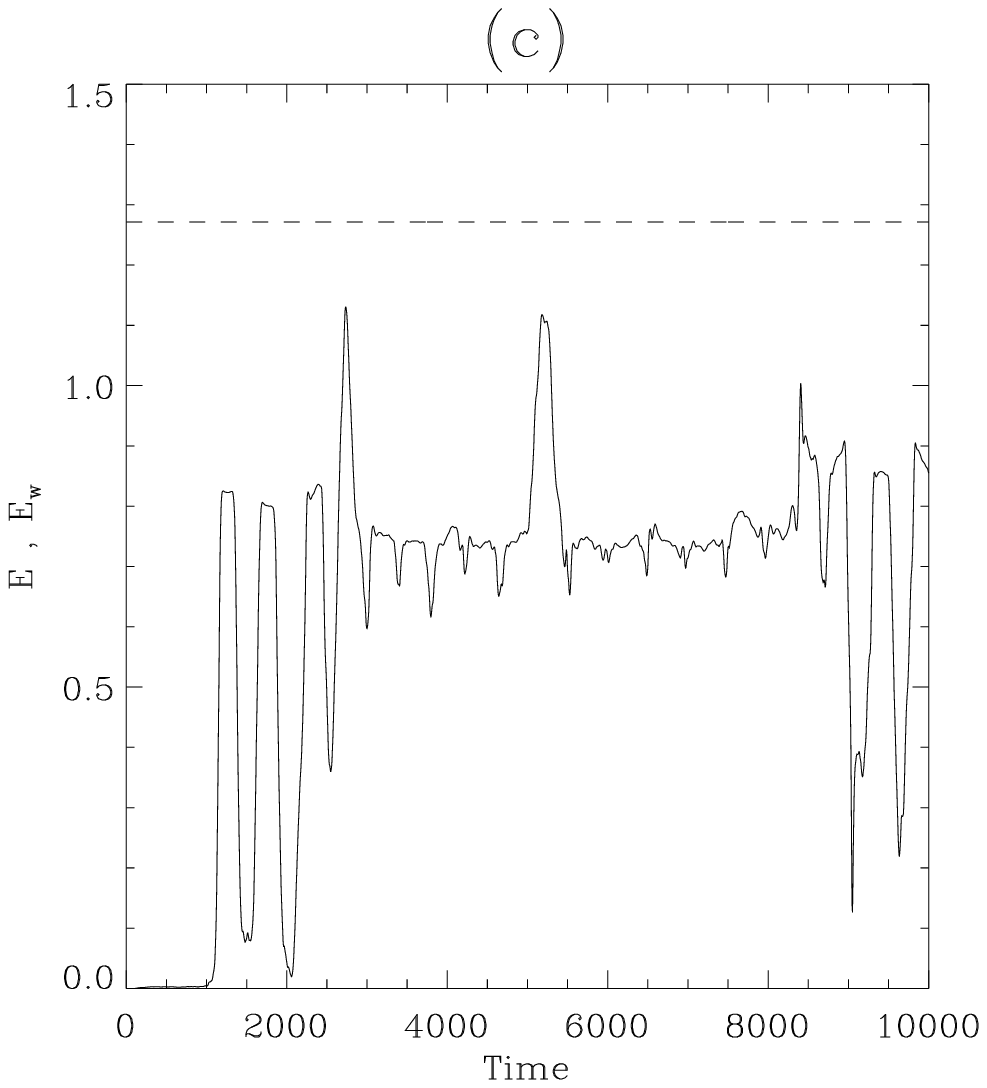}} \\
\end{center}
\caption{Interaction of two breathers with an extended defect. $K_n = 1.0$
outside the defect, $K_n = 0.5$ inside the defect which is 20 cells wide
(between cells 145 and 165 in a 200-cell lattice) The parameters
of the breathers are $\eta = 0.1$ ($u_e = -0.014, u_c = 0.333$)
and $\eta = 0.2$ ($u_e = -0.014, u_c = 1.356$), and both have
a carrier wavector $q=0.1$. (a) Three-dimensional
picture of the energy density $e_n$ versus time, and 
(b) contour plot of the same
energy density. (c) Energy $E_w$ inside the defect window (full line) 
and total energy of the chain $E$ (dotted line) versus time.}
\label{figu5}
\end{figure}

\begin{figure}[bth]
\begin{center}
\mbox{\psboxto(10.cm;10.cm){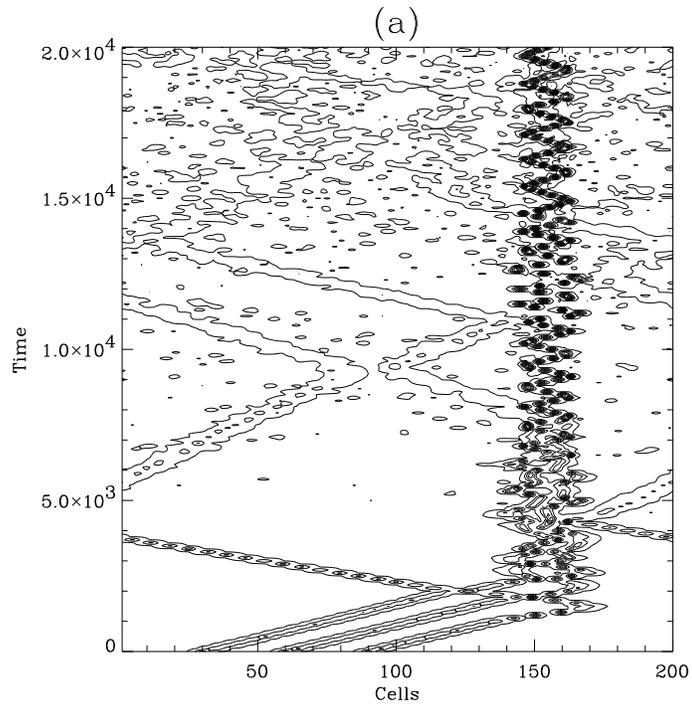}} \\
\mbox{\psboxto(10.cm;10.cm){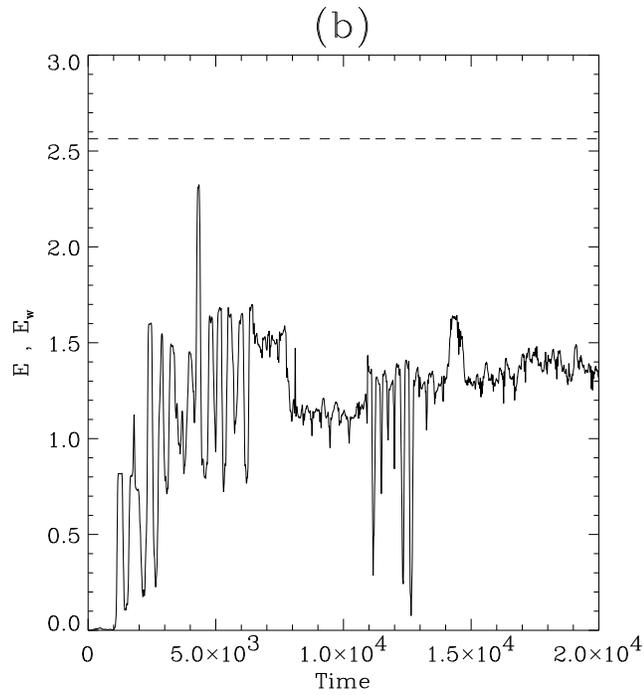}} \\
\end{center}
\caption{Interaction of three breathers with an amplitude $\eta = 0.2$ with
the same extended defect as in Fig.~5. (a) Contour plot of the energy density
in the chain versus time. (b)  Energy $E_w$ inside the defect window 
(full line)  and total energy of the chain $E$ (dotted line) versus time.}
\label{figu6}
\end{figure}

\begin{figure}[bth]
\begin{center}
\mbox{\psboxto(10.cm;10.cm){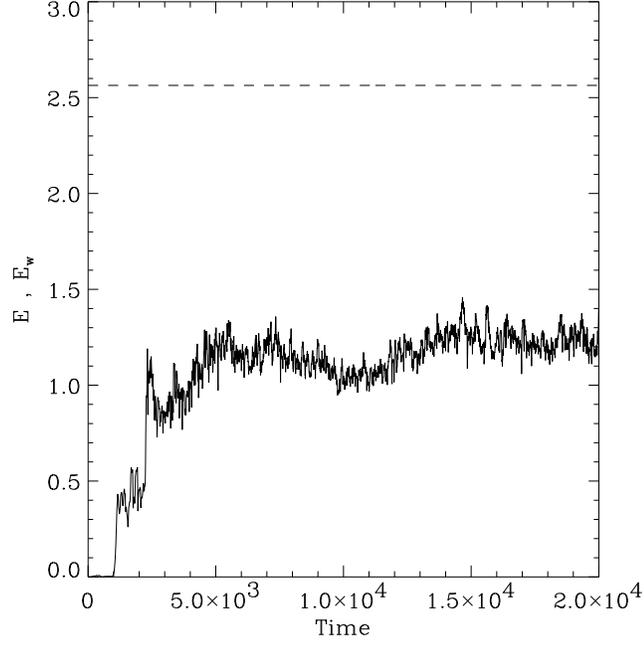}} \\
\end{center}
\caption{Energy in the defect region versus time for the interaction of
three breathers with an extended defect. The conditions are exactly the
same as in Fig.~6, but, in addition to changing locally the coupling
constant inside the defect, we also reduce the barrier of the Morse
potential by a factor 0.8 in this region to model some enzyme specific
action.}
\label{figu7}
\end{figure}


\begin{table}[p]
\begin{center}
\begin{tabular}{| c | l | c || l | c |}
             &           & $\kappa = -0.4$ &       & $\kappa = 0.4$ \\ 
\hline
             & passed    & $< 0.5$     & passed    & $< 0.5$ \\
$ \xi = 0.2$ & reflected & $0.6 - 1.1$ & reflected & $0.6 - 1.4$ \\
             & trapped   & $> 1.2$     & passed    & $> 1.5$ \\
\hline
$\xi = -0.2$ & reflected &  $< 1.1$ & reflected & all values \\
             & trapped   &  $> 1.2$ &        &   \\
\end{tabular}
\caption{Behavior of a breather determined by collective coordinate
method with initial values $\phi = -0.4$, $\zeta =-13$
and a two-step defect with length $l=10$.
The table lists the outcome of the interaction of the breather with the
defect for different values of $\eta$, and two different signs for
$\kappa$.}
\end{center}
\label{tab}
\end{table}

 \end{document}